\documentclass{elsart}

 \usepackage{graphicx}
\usepackage{amssymb}

\providecommand\boldsymbol[1]{\mbox{\boldmath $#1$}}
\providecommand\bnabla{\boldsymbol{\nabla}}
\providecommand\bcdot{\boldsymbol{\cdot}}

\newcommand\Mb{\mathbf{M}}
\newcommand\dr{\mathrm{d}}
\newcommand\Pb{\mathbf{P}}

\newcommand\ub{\mathbf{u}}
\newcommand\xb{\mathbf{x}}
\newcommand\kb{\mathbf{k}}

\newcommand\Ub{\boldsymbol{U}}

\newcommand\er{\mathrm{e}}
\newcommand\ir{\mathrm{i}}

\newcommand\hu{\widehat{u}}
\newcommand\hw{\widehat{w}}
\newcommand\uL{\overline{u}^L}

\newcommand\zL{\overline{\zeta}^L}
\newcommand\zb{\overline{\zeta}}

\begin{document}

\begin{frontmatter}



\title{Explicit wave-averaged primitive equations using a Generalized Lagrangian Mean}


\author[1]{Fabrice Ardhuin},
\author[1,2]{Nicolas Rascle},
\author[3]{K. A. Belibassakis}

\address[1]{Centre Militaire d'Oc{\'e}anographie, Service
Hydrographique et Oc\'{e}anographique de la Marine, 29609 Brest, France}
\address[2]{Laboratoire de Physique des Oc{\'e}ans, Universit{\'e} de Bretagne Occidentale, 29000 Brest, France}
\address[3]{School of Naval Architecture and Marine Engineering, National Technical
University of Athens, Athens, Greece}

\begin{abstract}
The generalized Langrangian mean theory provides exact equations
for general wave-turbulence-mean flow interactions in three
dimensions. For practical applications, these equations must be
closed by specifying the wave forcing terms. Here an approximate
closure is obtained under the hypotheses of small surface slope,
weak horizontal gradients of the water depth and mean current, and
weak curvature of the mean current profile. These assumptions
yield analytical expressions for the mean momentum and pressure
forcing terms that can be expressed in terms of the wave spectrum.
A vertical change of coordinate is then applied to obtain
$glm2z$-RANS equations (55) and (57) with non-divergent mass
transport in cartesian coordinates. To lowest order, agreement is
found with Eulerian-mean theories, and the present approximation
provides an explicit extension of known wave-averaged equations to
short-scale variations of the wave field, and vertically varying
currents only limited to weak or localized profile curvatures.
Further, the underlying exact equations provide a natural
framework for extensions to finite wave amplitudes and any
realistic situation. The accuracy of the approximations is
discussed using comparisons with exact numerical solutions for
linear waves over arbitrary bottom slopes, for which the equations
are still exact when properly accounting for partial standing
waves.  For finite amplitude waves it is found that the
approximate solutions are probably accurate for ocean mixed layer
modelling and shoaling waves, provided that an adequate turbulent
closure is designed. However, for surf zone applications the
approximations are expected to give only qualitative results due
to the large influence of wave nonlinearity on the vertical
profiles of wave forcing terms.
\end{abstract}

\begin{keyword}
radiation stresses \sep Generalized Lagrangian Mean \sep
wave-current coupling \sep drift \sep surface waves \sep three
dimensions

 \end{keyword}
\end{frontmatter}

\section{Introduction}
From wave-induced mixing and enhanced air-sea interactions in deep water, to
wave-induced currents and sea level changes on beaches, the effects of waves on
ocean currents and turbulence are well documented (e.g. Battjes 1988, Terray et
al. 1996\nocite{Battjes1988,Terray&al.1996}). The refraction of waves over
horizontally varying currents is also well known, and the modifications of
waves by vertical current shears have been the topic of a number of theoretical
and laboratory investigations (e.g. Biesel 1950, Peregrine
1976\nocite{Peregrine1976}, Kirby and Chen
1989\nocite{Biesel1950,Kirby&Chen1989}, Swan et al. 2001\nocite{Swan&al.2001}),
and field observations (e.g. Ivonin et al. 2004\nocite{Ivonin&al.2004}). In
spite of this knowledge and the importance of the topic for engineering and
scientific applications, ranging from navigation safety to search and rescue,
beach erosion, and de-biasing of remote sensing measurements, there is no well
established and generally practical numerical model for wave-current
interactions in three dimensions.

Indeed the problem is made difficult by the difference in time
scales between gravity waves and other motions. When motions on
the scale of the wave period can be resolved, Boussinesq
approximation of nearshore flows has provided remarkable numerical
solutions of wave-current interaction processes (e.g. Chen et al.
2003\nocite{Chen&al.2003}, Terrile et al.
2006\nocite{Terrile&al.2006}). However, such an approach still
misses some of the important dynamical effects as it cannot
represent real vertical current shears and their mixing effects
(Putrevu and Svendsen 1999\nocite{Putrevu&Svendsen1999}). This
shortcoming has been partly corrected in quasi-three dimensional
models (e.g. Haas et al. 2003\nocite{Haas&al.2003}), or
multi-layer Boussinesq models (e.g. Lynnett and Liu
2005\nocite{Lynett&Liu2005}).

The alternative is of course to use fully three dimensional (3D)
models, based on the primitive equations. These models are
extensively used for investigating the global, regional or coastal
ocean circulation (e.g. Bleck 2002\nocite{Bleck2002}, Shchepetkin
and McWilliams 2003\nocite{Shchepetkin&McWilliams2003}). An
average over the wave phase or period is most useful due to
practical constraints on the computational resources, allowing
larger time steps and avoiding non-hydrostatic mean flows.
Wave-averaging also allows an easier interpretation of the model
result. A summary of wave-averaged models in 2 or 3 dimensions is
provided in table
1\nocite{Jenkins1987}\nocite{Newberger&Allen2007a}.
\begin{table}
  \centering
  \begin{tabular}{cccccc}
\hline
  Theory          &  averaging  & momentum variable & main limitations  \\
\hline
  Phillips (1977) &   Eulerian                 & total ($U$)   & 2D, $d\overline{u}/dz=0$   \\
  Garrett (1976) & Eulerian                  & mean flow ($U-M^w/D$)   & 2D, $d\overline{u}/dz=0$, $kh \gg 1$   \\
  Smith (2006) &  Eulerian                  & mean flow  ($U-M^w/D$)  & 2D, $d\overline{u}/dz=0$   \\
  GLM (A\&M 1978a) &  GLM  & mean flow ($\overline{\mathbf u}^L-\mathbf{P}$) &  none (exact theory) \\
  aGLM (A\&M 1978a) &     GLM  & total ($\overline{\mathbf u}^L$) &  none (exact theory) \\
    Leibovich (1980) &   Eulerian  & mean flow ($\overline{\mathbf u}^L-\mathbf{P}$) &  2nd order, $\nu$ constant  \\
  Jenkins (1987) &   GLM  &  mean flow ($\overline{\mathbf u}^L-\mathbf{P}$) &  2nd order, horizontal uniformity   \\
  Groeneweg (1999) &   GLM  & total ($\overline{\mathbf u}^L$) &  2nd order  \\
  Mellor (2003) &   following $\xi_3$ & total ($\overline{\mathbf u}^L$) &  2nd order, flat bottom  \\
  MRL04 &   Eulerian & mean flow ($\overline{u}$) &  below troughs, $\overline{u}  \ll C$, $\nu=0$  \\
  NA07 &   Eulerian & mean flow ($\overline{u}$) &  below troughs, 2nd order, $kH \ll 1$  \\
  present paper  & GLM  &  mean flow ($\overline{\mathbf u}^L-\mathbf{P}$) &  2nd order  \\
\hline
\end{tabular}
  \caption{Essential attributes of some general wave-current coupling theories. See list of symbols for details (table 2 at the end of the paper).
  Although Mellor (2003) derived his wave-averaged equations
  with spatially varying wave amplitudes, his use of flat-bottom Airy wave kinematics is inconsistent with the presence of bottom slopes (see ARB07).
  MRL04 stands for McWilliams et al. (2004) and NA2007 stands for Newberger and Allen (2007).}\label{table_theory}
\end{table}

\subsection{Air-water separation}
\begin{figure}[htb]
 \vspace{9pt}
\centerline{\includegraphics[width=0.8\textwidth]{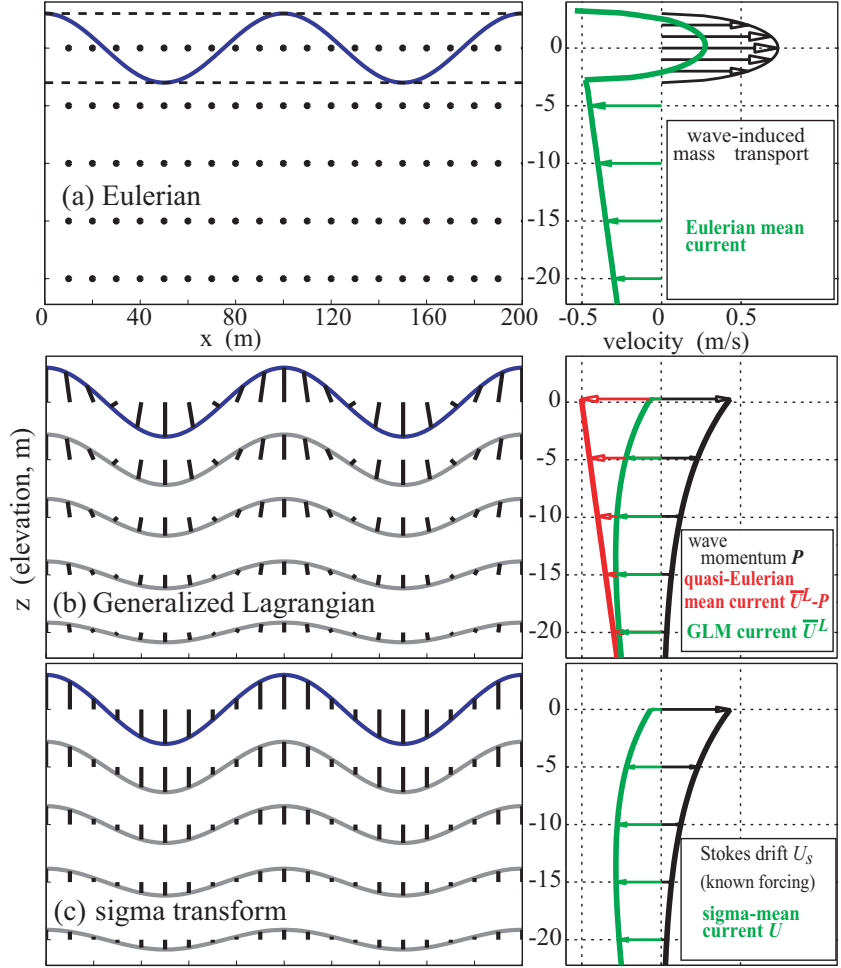}}
 \caption{Averaging procedures (left) and examples of resulting velocity profiles (right) in the
 case of (a) Eulerian averages (e.g. Rivero and Sanchez-Arcilla 1995, McWilliams et al. 2004),
  (b) the Generalized Lagrangian Mean (Andrews and McIntyre
  1978a), and (c) sigma transform (Mellor 2003, AJB07).
 The thick black bars connect the fixed points $\xb$ where the average field is evaluated, to the displaced
 points $\xb+\xi$ where the instantaneous field is evaluated. For averages in moving coordinates the
 points $\xb+\xi$ at a given vertical level $\xi$ are along the gray lines. The drift velocity is the sum
 of the (quasi-Eulerian) current and the wave-induced mass transport.
 In the present illustration an Airy wave
 of amplitude 3~m and wavelength 100~m in 30~m depth, is superimposed on a hypothetical current of velocity
 $u(z)=-0.5-0.01 z$~m/s for
 all $z< \zeta(\xb)$. The
 current profile is not represented in (c) since it is not directly given in Mellor's theory, although it can obviously be obtained by taking the
 difference of the other two profiles.}
 \label{figavg1}
\end{figure}

 In 3D, problems arise due to the presence of both air and water in
the region between wave crests and troughs. Various approaches to
the phase or time averaging of flow properties are illustrated in
figure \ref{figavg1} (see also Ardhuin et al. 2007b, hereinafter
ARB2007). For small amplitude waves, one may simply take a Taylor
expansion of mean flow properties (e.g. McWilliams et al.
2004\nocite{McWilliams&al.2004}, hereinafter MRL04). Using a
decomposition of the non-linear advection term in the equations of
motion $\ub \bcdot \bnabla \ub = \bnabla u^2 + \ub \times \bnabla
\ub$, McWilliams et al. (2004, see also Lane et al. 2007) obtained
a relatively simple set of equation for conservative wave motion
over sheared currents, for a given choice of small parameters.
These parameters include the surface slope $\varepsilon_1=k_0 a_0$
and the ratio of the wavelength and scale of evolution of the wave
amplitucde. Further, these equations were derived with a scaling
corresponding to a non-dimensional depth $k_0 h_0$ of order 1,
with $k_0$, $a_0$ and $h_0$ typical values of the wavenumber, wave
amplitude and water depth, respectively. These authors also
assumed that the current velocity was of the same order as the
wave orbital velocity, both weaker than the phase speed by a
factor $\varepsilon_1$. That latter assumption may generally be
relaxed since the equations of motion are invariant by a change of
reference frame, so that only the current vertical shear may need
to be small compared to the wave radian frequency, provided that
the current, water depth and wave amplitudes are slowly varying
horizontally.


For waves of finite amplitude, a proper separation of air and
water in the averaged equations of motion requires a change of
coordinates that maps the moving free surface to a level that is
fixed, or at least slowly varying. This is usual practice in
air-sea interaction studies, and it has provided approximate
solutions to problems such as wind-wave generation or
wave-turbulence interactions (e.g. Jenkins 1986, Teixeira and
Belcher 2002\nocite{Jenkins1986,Teixeira&Belcher2002}) but it
brings some complications. The most simple change of coordinate
was recently proposed by Mellor (2003\nocite{Mellor2003}), but it
appears to be impractical in the presence of a bottom slope
because its accurate implementation requires the wave kinematics
to first order in the wave slope (Ardhuin et al., 2007b,
hereinafter ARB07\nocite{Ardhuin&al.2007b,Rivero&Arcilla1995}).

\subsection{Separation of wave and current momentum fluxes}
Another approach is to use one of the two sets of exact averaged equations
derived by Andrews and McIntyre (1978a\nocite{Andrews&McIntyre1978a}).
Groeneweg (1999) successfully used the second set, the alternative Generalized
Lagragian Mean equations (aGLM), approximated to second order in wave slope,
for the investigation of current profile modifications induced by waves (see
also Groeneweg and Klopman 1998, Groeneweg and Battjes
2003\nocite{Groeneweg&Klopman1998,Groeneweg1999,Groeneweg&Battjes2003}). This
work was also loosely adapted for engineering use in the numerical model
Delft3D (Walstra et al. 2001\nocite{Walstra&al.2001}).

However, aGLM equations describe the evolution of the total flow
momentum, which includes the wave pseudo-momentum per unit mass
$\Pb$. That vector quantity is generally close to the Lagrangian
Stokes drift $\overline{\mathbf u}^S$ (see below), and it is not
mixed by turbulence\footnote{The Stokes drift is a residual
velocity over the wave cycle, its mixing is not possible without a
profound modification of the wave kinematics.}, unlike the mean
flow momentum. Further, $\Pb$ is carried by the wave field at the
group velocity, which is typically one order of magnitude faster
than the drift velocity. Thus bundling $\Pb$ with the rest of the
momentum may lead to large errors with the turbulence closure.
Other practical problems arise due to the strong surface shear of
$\Pb$ and $\overline{\mathbf u}^S$ (e.g. Rascle et al.
2006\nocite{Rascle&al.2006}) whereas the quasi-Eulerian current is
relatively uniform in deep water (e.g. Santala and Terray
1992\nocite{Santala&Terray1992}). Thus solving for the total
momentum (including $\Pb$) requires a high resoltion near the
surface. Finally, a consistent expression of the aGLM equations
with a sloping bottom and wave field gradients is difficult due to
the divergence of vertical fluxes of momentum (vertical radiation
stresses) that must be expressed to first order in all the small
parameters that represent the slow wave field evolution (bottom
slope, wave energy gradients, current shears...). This same
problem arises with Mellor's (2003) equations and is discussed in
ARB07.

The first set of GLM equations describes the evolution of the
quasi-Eulerian current only, and, just like the decomposition of
$\ub \bcdot \bnabla u$ used by MRL04, it does not require the
evaluation of these vertical radiation stresses. These equations
were used by Leibovich (1980) to derive the Craik-Leibovich
equations that is the basis of theories for Langmuir circulations.
 However, in that work he did not attempt an explicit integration of the GLM set,
 and thus did not express the wave forcing terms from wave amplitudes or spectra.
 The general mathematical structure of the GLM equations and their conservatin properties are also
 well detailed in Holm (2002) and references therein.

 Further, the GLM flow is generally divergent as the averaging
operator introduces an implicit change of the vertical coordinate.
This question has been largely overlooked by previous users of GLM
theory (Leibovich 1980\nocite{Leibovich1980}, Groeneweg 1999).
Further, in order to be implemented in a numerical model, the
wave-induced forcing terms must be made explicit using approximate
solutions for wave-induced motions and pressure. We will assume
that the slowly varying spectrum is known, typically provided by a
wave model. Given the degree of accuracy attained by modelled wave
spectra in a wide variety of conditions this is generally
appropriate (e.g. Herbers et al. 2000, Ardhuin et al. 2003, 2007,
Magne et al.
2007\nocite{Herbers&al.2000,Ardhuin&al.2003b,Ardhuin&al.2007a,Magne&al.2007}).
We note in passing that no explicit and theoretically satisfying
theory is available for the transport of the wave action spectrum
over vertically and horizontally sheared currents. Indeed, the
exact theory of Andrews and McIntyre
(1978b\nocite{Andrews&McIntyre1978b}) is implicit and would
require an explicit approximation of the wave action from know
wave kinematics, similar to the approximation of the wave
pseudo-momentum performed here.

The goal of the present paper is to provide a practical and accurate method for
wave-current coupling that is general enough for applications ranging from the
ocean mixed layer to, possibly, the surf zone. GLM equations, for the reasons
listed above, are a good candidate for this application. Although not as simple
as an
 Eulerian average, the GLM operator is capable of properly separating air
and water in the crest to trough region, leading to physically
understandable definitions of mean properties on either side of
the air-sea interface. The practical use of GLM requires some
approximations and transformations. We provide in section 2 a
derivation of explicit and approximate $glm2z$-RANS equations.
Given the large literature on the subject, we explore in section 3
the relationships between GLM, aGLM and other forms of
wave-averaged 3D and depth-integrated 2D equations. A preliminary
analysis of the expected errors due to the approximations are
provided in section 4, and conclusions follow in section 5. Full
numerical solutions using the $glm2z$-RANS equations will be
reported elsewhere, in particular in the doctorate thesis of
Nicolas Rascle.

\section{glm2-RANS equations}
\subsection{Generalities on GLM and linear wave kinematics} We first define the
Eulerian average $\overline{\phi\left({\mathbf x},t\right)}$ of
$\phi\left({\mathbf x},t\right)$, where the average may be an average over
phase, realizations, time $t$ or space. We now take this average at displaced
positions $\mathbf{x}+\xi$, with $\xi=(\xi_1,\xi_2,\xi_3)$ a displacement
vector, and we defining the velocity ${\mathbf v}$ at which the mean position
is displaced when the actual position moves at the fluid velocity
$\ub(\xb+\xi)$.  One obtains the corresponding GLM of $\phi$
\begin{equation}
\overline{\phi(\mathbf{x},t)}^L = \overline{\phi(\mathbf{x}+\xi,t)}\label{phiL}
\end{equation}
by choosing the displacement field $\xi$ so that
\begin{itemize}
  \item the mapping $ \mathbf{x} \rightarrow \mathbf{x}+\xi$ is invertible
  \item $\overline{\xi\left({\mathbf
x},t\right)}=0$
  \item $\overline{{\mathbf v}\left({\mathbf x},t
\right)}={\mathbf v}\left({\mathbf x},t\right)$, which gives ${\mathbf
v}=\overline{\ub(\mathbf{x},t)}^L$.
\end{itemize}
Such a mapping is illustrated in figure 1.c for linear waves. Lagrangian
perturbations are logically defined as the field minus its average, i.e.,
\begin{equation}
\overline{\phi(\mathbf{x},t)}^l
=\phi(\mathbf{x}+\xi,t)-\overline{\phi(\mathbf{x},t)}^L=\phi(\mathbf{x}+\xi,t)-\overline{\phi(\mathbf{x}+\xi,t)}.\label{Lagr.pert}
\end{equation}
Here we shall take our Eulerian average to be a phase average\footnote{For
uncorrelated wave components the phase average is obtained by the sum of the
phase averages of each component. In the presence of phase correlations, such
as in the case of partially standing waves or nonlinear phase couplings, the
sum has to be averaged in a coherent manner.}. Given any Eulerian flow field
$\ub(\xb,t)$, one may define a first displacement by
\begin{equation}
\xi'(\xb,t,\Delta t)=\int_t^{t+\Delta t} \ub(\xb+\xi'(\xb,t,t'-t),t')\dr
t'.\end{equation}  The mean drift velocity is defined as  ${\mathbf
v}(\xb,t)=\lim_{\Delta t\rightarrow 0} \overline{\xi'(\xb,t,\Delta t)}/(\Delta
t)$. The GLM displacement field is then given by $\xi=\xi'- {\mathbf v} t -
\overline{\xi'- {\mathbf v} t}$. This construction of ${\mathbf v}$ and $\xi$
guarantees that the required properties are obtained, provided that the limit
$\Delta t \rightarrow 0$ commutes with the averaging operator. For periodic
motions one may also take ${\mathbf
v}=\overline{\left(\xi'(t+T^L)-\xi'(t)\right)}/(T^L)$, with $T^L$ the
Lagrangian wave period (the time taken by a water particle to return to the
same wave phase). This definition will be used for Miche waves in section 4.2.

Clearly GLM differs from the Eulerian mean. The difference between the two is
given by the Stokes correction (Andrews et McIntyre 1978a). Below the wave
troughs, the Stokes correction for the velocity is the Stokes drift, by
definition,
\begin{equation}
\overline{\mathbf u}^S \equiv \overline{{\mathbf
u}}^L-\overline{\ub}.\label{Us}
\end{equation}
More generally, for a continuously differentiable field $\phi$ the Stokes
correction is given by (Andrews and McIntyre 1978a, equation 2.27),
\begin{equation}
\overline{\phi}^L \equiv
\overline{\phi}+\overline{\phi}^S=\overline{\phi}+\overline{\xi_j
\frac{\partial \phi}{\partial x_j}}+\frac{1}{2}\overline{\xi_j
\xi_k}\frac{\partial^2 \overline{\phi}}{\partial x_j \partial
x_k}+O\left(\max_{i,j,k} \left\{\frac{\partial^3
\overline{\phi}}{\partial x_i
\partial x_j  \partial x_k}\right\} |\xi|^3\right), \label{Stokes correction}
\end{equation}
with an implicit summation over repeated indices.

The GLM average commutes with the Lagrangian derivative, thus the GLM velocity
$\overline{{\mathbf u}}^L$ is the average drift velocity of water particles.
One should however be careful that the GLM average does not commute with most
differential operators, for example the curl operator. Indeed the GLM velocity
of irrotational waves is rotational, which is clearly apparent in the vertical
shear of the Stokes drift (see also Ardhuin and Jenkins
2006\nocite{Ardhuin&Jenkins2006} for a calculation of the lowest order mean
shears $\overline{\partial u_\alpha /\partial z}^L$ and $\overline{\partial u_3
/\partial x}^L$).

One of the interesting aspects of GLM theory is that it clearly separates the
wave pseudo-momentum $\Pb$ from the quasi-Eulerian mean momentum
$\widehat{{\mathbf u}}=\overline{{\mathbf u}}^L-\Pb$. This is a key aspect for
numerical modelling since $\Pb$ is transported by the wave field at the group
velocity, of the order of 5~m~s$^{-1}$ in deep water, while $\widehat{{\mathbf
u}}$ is transported at the much slower velocity $\overline{{\mathbf u}}^L$.
$\Pb$ is defined by (Andrews and McIntyre 1978a, eq.
 3.1),
\begin{equation}
P_i=-\overline{\xi_{j,i} \left(u^l_j + \epsilon_{j k l} f_k \xi_l /2
\right)}\label{Prot},
\end{equation}
where $\epsilon_{i j k} A_j B_k$ is the $i$-component of the
vector product $\mathbf{A} \times \mathbf{B}$, and $f_k/2$ is the
$k$-component of the rotation vector of the reference frame. In
the applications considered here the effect of rotation can be
neglected in (\ref{Prot}) due to the much larger rotation period
of the Earth compared to the wave period. We will thus take
\begin{equation}
P_i=-\overline{\xi_{j,i} u^l_j}\label{P}.
\end{equation}

For practical use, the GLM equations have to be closed by
specifying the wave-induced forcing terms. In order to give
explicit approximations for the wave-induced effects, we will
approximate the wave motion as a sum of linear wave modes, each
with a local wave phase $\psi$ giving the local wave number
$\kb=(k_1,k_2)=\bnabla \psi$, and radian frequency
$\omega=-\partial \psi / \partial t$, and an intrinsic linear wave
radian frequency $\sigma=\left[g k \tanh (kD)\right]^{1/2}=\omega
- \kb \bcdot \Ub_A$, where $\Ub_A$ is the phase advection
velocity, $D$ is the local mean water depth, and $g$ the
acceleration due to gravity and Earth rotation. Defining
$h(x_1,x_2)$ as the local depth of the bottom and
$\zeta(x_1,x_2,t)$ as the free surface elevation, one has
$D=\overline{\zeta}+h$. We assume that the wave slope
$\varepsilon_1 =\max\left(|\bnabla \zeta|\right)$ is small
compared to unity (this will be our first hypothesis H1), with
$\bnabla$ denoting the horizontal gradient operator. We also
restrict our investigations to cases for which the Ursell number
is small $Ur=(a/D)/(kD)^2 < 1$ (this is hypothesis H2). We further
restrict our derivations to first order in the slow spatial scale
$\varepsilon_2$. That small parameter may be defined as the
maximum of the slow spatial scales $|(\partial a/\partial x)/(k
a)|$, $|(\partial \overline{u}/\partial x)/(\sigma)|$, $|(\partial
D/\partial x)|$, and time scales $|(\partial a/\partial t)/(\sigma
a)|$, $|k (\partial \hu/\partial t)/(\sigma)^2$, and $|(\partial
D/\partial t)k/\sigma|$ (hypothesis H3). It will also appear that
the current profile may cause some difficulties. Since we have
already assumed a small wave steepness we may use Kirby and Chen's
(1989) results, giving the dispersion relation
\begin{equation}
\omega=\sigma + k_\alpha \int_{-h}^{\zb} \hu_\alpha \frac{2  k
\cosh\left[2k(z+h)\right]}{\sinh(2kD)} \dr z +
O(\varepsilon_3),\label{dispersion}
\end{equation}
where $\alpha$ is a dummy index representing any horizontal component 1 or 2,
and the summation is implicit over repeated indices. The index 3 will represent
the vertical components positive upwards, along the direction $z=x_3$. In
particular we shall assume that their correction to the lowest order stream
function (their eq. 23) is relatively small, which may be obtained by requiring
that the curvature of the current is weak or concentrated in a thin boundary
layer, i.e. $\varepsilon_3 \ll 1$ (hypothesis H4) with
\begin{equation}
\varepsilon_3=\frac{1}{\omega \sinh(kD)} \int_{-h}^\zeta \left|\frac{\partial^2
\overline{u}}{
\partial z^2}\right| \sinh \left[2 k (z+h)\right] \dr z.\label{eps3}
\end{equation}
For simplicity we will further require that $a^2 \left[{\partial^3
\overline{u}_\alpha}/{\partial z^3}/({\sigma}) \right] \leq
\varepsilon_3 $ (hypothesis H5), which may be more restrictive
than H4. Finally, we will neglect the vertical velocity
$\widehat{w}$ in the vertical momentum equation for the mean flow
momentum  (i.e. we assume the mean flow to be hydrostatic, this is
our hypothesis H6).

In the following we take $\varepsilon=\max{\varepsilon_i , 1 \leqslant i
\leqslant 3}$. The wave-induced pressure and velocity are given by
\begin{eqnarray}
\widetilde{p}&=& \rho_w g a \left[ F_{CC} \cos\psi + O(\varepsilon)\right] \label{p}\\
\widetilde{u}_\alpha& =& a \sigma \frac{k_\alpha}{k} \left[ F_{CS} \cos\psi + O(\varepsilon)\right] \label{u1}\\
\widetilde{u}_3& =& a \sigma \left[ F_{SS} \sin\psi +
O(\varepsilon)\right], \label{w}
\end{eqnarray}
where $a$ is the local wave amplitude, $\rho_w$ is the water density, taken
constant in the present paper. We have used the short-hand notations
$F_{CC}=\cosh(kz+kD)/\cosh(kD)$, $F_{CS}=\cosh(kz+kD)/\sinh(kD)$, and
$F_{SS}=\sinh(kz+kD)/\sinh(kD)$.

From now on, only the lowest order approximations will be given unless
explicitly stated otherwise. In order to estimate quantities at displaced
positions, the zero-mean displacement field is given by
\begin{eqnarray}
u^l_i & \equiv & \ub(\xb+\xi)-\uL_i  \nonumber \\
&\simeq &\widetilde{u}_i + \xi_j \frac{\partial \overline{u}_i}{\partial
x_j}+\left(\xi_j \frac{\partial \widetilde{u}_i}{\partial x_j}-\overline{\xi_j
\frac{\partial \widetilde{u}_i}{\partial x_j}}\right)+
\frac{1}{2}\left(\xi_j^2-\overline{\xi_j^2}\right)\frac{\partial^2
\overline{u}_i}{\partial x_j^2}. \label{Lagr. pert.2}
\end{eqnarray}
Thanks to the definition of $\uL$, we also have
\begin{equation}
u^l_i =\frac{\partial \xi_i}{\partial t}+\uL_j \frac{\partial
\xi_i}{\partial x_j} \simeq  \frac{\partial \xi_i}{\partial
t}+\uL_\alpha \frac{\partial \xi_i}{\partial x_\alpha},
\label{Lagr. pert}
\end{equation}
in which the vertical velocity has been neglected. The greek
indices $\alpha$ and $\beta$ stand for horizontal components only.

To lowest order in the wave amplitude, the displacements $\xi_i$
and Lagrangian velocity perturbations $u^l_i$ are obtained from
(\ref{Lagr. pert.2}) and (\ref{Lagr. pert}),
\begin{eqnarray}
u_3^l& =& \widetilde{u}_3  \label{ul3} \\
{\xi_3}& =& a m \left[F_{SS} \cos\psi\right] \label{xi3}  \\
u_\alpha^l& =& \widetilde{u}_\alpha + {\xi_3} \frac{\partial
\overline{u}_\alpha}{\partial z} +{\xi_\beta} \frac{\partial
\overline{u}_\alpha}{\partial x_\beta} +  O\left(\sigma k
a^2\right) \cos 2\psi
 +O\left(a^3
\frac{\partial^2
\overline{u}_\alpha}{\partial z^3}\right) \label{ulalpha}\\
&\simeq & a \left[ \sigma \frac{k_\alpha}{k} F_{CS} + m F_{SS}
\frac{\partial
\overline{u}_\alpha}{\partial z} \right] \cos\psi \label{ulalpha2} \\
{\xi_\alpha}& =& -a m \left[\frac{k_\alpha}{k} F_{CS} + \frac{m}{\sigma}
\frac{\partial \overline{u}_\alpha}{\partial z} F_{SS} \right] \sin\psi +
O\left(\frac{a^2}{\sigma} \frac{\partial^2 \overline{u}_\alpha}{\partial
z^2}\right)\sin 2 \psi \nonumber \\
& &  + O\left(\frac{a}{\sigma} \frac{\partial \overline{u}_\alpha}{\partial
x_\beta}\right) \cos\psi + O\left(\frac{a^3}{\sigma} \frac{\partial^2
\overline{u}_\alpha}{\partial z^3}\right),\label{xi1}
\end{eqnarray}
The shear correction parameter $m$, arising from the time-integration of
(\ref{Lagr. pert}), is given by
\begin{equation}
m(\xb,\kb,z,t)=\frac{\sigma}{\omega-\kb \bcdot
\overline{\ub}^L(\xb,z,t)}.\label{m}
\end{equation}
Based on (\ref{dispersion}) $m$ differs from 1 by a quantity of order
$\sigma^{-1} \partial \overline{u}/\partial z$.

Using our assumption (H5) the last term in eq. (\ref{xi1})  may be
neglected. The last two term in eq. (\ref{ulalpha}) have been
neglected  because they will give negligible $O(\varepsilon^3)$
terms in $\Pb$, $\zL$ or other wave-related quantities, when
multiplied by other zero-mean wave quantities.

Using the approximate wave-induced motions, one may estimate the Stokes drift
\begin{eqnarray}
    \overline{\mathbf u}^S& \equiv &\overline{{\mathbf u}}^L-\overline{u}\simeq \overline{\xi \bcdot \bnabla \widetilde{u}}+\frac{1}{2} \overline{\xi_3^2}
\frac{\partial^2 \overline{u}_\alpha}{\partial z^2}\nonumber \\
& =&\frac{m a^2}{4 \sinh^2(kD)}\left[2 \sigma {\mathbf k}
\cosh(2kz+2kh)+{\mathbf k} m \sinh(2kz+2kh) \frac{\kb}{k} \bcdot \frac{\partial
\overline{\ub}}{\partial
z}\right. \nonumber \\
 & &+\left.\frac{\partial^2 \overline{\ub}}{\partial z^2}
\sinh^2(kz+kh)\right],\label{Us_approx}
\end{eqnarray}
the horizontal wave pseudo-momentum
\begin{eqnarray}
    P_\alpha&=&-\overline{\frac{\partial \xi_\beta}{\partial x_\alpha} u_\beta^l} -\overline{\frac{\partial \xi_3}{\partial x_\alpha} w^l}
    \nonumber \\
& \simeq&\frac{m a^2}{4 \sinh^2(kD)}\left[2 \sigma k_\alpha
\cosh(2kz+2kh)+2 k_\alpha  m \sinh(2kz+2kh) \frac{k_\alpha}{k}
\bcdot \frac{\partial \overline{\ub}}{\partial
z}\right. \nonumber \\
 & &+2 m^2 \frac{k_\alpha}{\sigma} \sinh^2(kz+kh)\left.\left(\frac{\partial\overline{\ub}}{\partial z}
\right)^2\right],\label{Papprox}
\end{eqnarray}
and the GLM position of the free surface
\begin{equation}
   \zL=\zb +\zb^S=\zb
   +  \frac{\partial \zeta}{\partial x_\alpha}\xi_\alpha|_{z=\zb} =\zb + \frac{m a^2}{2} \left[\frac{k}{\tanh
   kD}+\frac{m \kb}{\sigma} \bcdot \frac{\partial \overline{\ub}}{\partial z}|_{z=\zb}\right].\label{zetaL}
\end{equation}
Thus the GLM of vertical positions in the water is generally
larger than the Eulerian mean of the position of the same
particles (see also McIntyre 1988\nocite{McIntyre1988}). This is
easily understood, given that there are more particles under the
crests than under the troughs (figure \ref{figavg1}.c). As a
result, the original GLM equations are divergent ($\bnabla \bcdot
\overline{u}^L \neq 0$) and require a coordinate transformation to
yield a non-divergent velocity field. That transformation is
small, leading to a relative correction of order
$\varepsilon_1^2$. That transformed set of equation is a modified
primitive equation that may be implemented in existing ocean
circulation models.

The horizontal component of the wave pseudo-momentum  $P_\alpha$
differs from the Stokes drift $\overline{u}_{\alpha}^S$ due to the
current vertical shear. Therefore the quasi-Eulerian mean velocity
$\hu_\alpha=\uL_\alpha-P_\alpha$ also differs from the Eulerian
mean velocity
$\overline{u}_\alpha=\uL_\alpha-\overline{u}_{\alpha}^S$
\begin{equation}
\hu_\alpha=\overline{u}_\alpha + \frac{1}{2} \overline{\xi_3^2}
\frac{\partial^2 \overline{u}_\alpha}{\partial z^2} +O(\varepsilon_3).
\label{Eul-quasi}
\end{equation}

The vertical wave pseudo-momentum $P_3=0$ is, at most, of order
$\sigma \varepsilon^3/k$. Although it may be neglected in the
momentum equation, it plays an important role in the mass
conservation equation, and will thus be estimated from $P_\alpha$.
In particular, for $m=1$ and in the limit of small surface slopes,
it is straightforward using (\ref{P}) to prove that $\Pb$ is
non-divergent and such that $\Pb \bcdot \mathbf{n}=0$ at $z=-h$,
with $\mathbf{n}$ the normal to the bottom. This gives,
\begin{equation}
P_3=- P_\alpha(-h) \frac{\partial h}{\partial x_\alpha} -
\int_{-h}^{z} \frac{\partial P_\alpha(z')}{\partial x_\alpha}
\mathrm{d}z' \label{P3}.
\end{equation}

Although this equality is not obvious for $m \neq 1$ and nonlinear
waves, corrections to (\ref{P3}) are expected to be only of higher
order ,in particular once $\Pb$ is transformed to $z$ coordinates.
Indeed, in the absence of a mean flow
$\Pb=\overline{\mathbf{u}}^L$ and it is non-divergent (see section
2.1.1).

\subsubsection{\textit{glm}2-RANS equations}
The velocity field is assumed to have a unique decomposition in
mean, wave and turbulent components
$\ub=\overline{\ub}+\widetilde{\ub}+\ub^\prime$, with
$\left<{\ub^\prime}\right>=0$, the average over the flow
realizations for prescribed wave phases. The turbulence will be
assumed weak enough so that its effect on the sea surface position
is negligible. We note ${\mathbf X}$ the divergence of the
Reynolds stresses, i.e. $X_i=\partial \left<u^\prime_i
u^\prime_j\right> / \partial x_j$, and we apply the GLM average to
the Reynolds-Average Navier-Stokes equations (RANS). We shall now
seek an approximation to the GLM momentum equations by retaining
all terms of order $\rho_w g \varepsilon^3$ and larger in the
horizontal momentum equation, and all terms of order $\rho_w g
\varepsilon^2$ in the vertical momentum equation. The resulting
equations, that may be called the "\textit{glm}2-RANS" equations,
are thus more limited in terms of wave nonlinearity than the
Eulerian mean equations of MRL04. At the same time, random waves
are considered here and that the mean current may be larger than
the wave orbital velocity. Indeed we make no hypothesis on the
current magnitude, but only on the horizontal current gradients
and on the curvature of the current profile. The present
derivation differs from that of Groeneweg (1999) by the fact that
we use the GLM instead of the aGLM equations (see table 1). The
name for these equations is loosely borrowed from Holm
(2002\nocite{Holm2002}) who instead derived an approximate
Lagrangian to obtain the momentum equation, and did not include
turbulence.

In order to simplify our calculations we shall use the form of the
GLM equations given by Dingemans (1997\nocite{Dingemans1997a}, eq.
2.596) with $\rho_w$ constant, which, among other things, removes
terms related to the fluid thermodynamics. The evolution equation
for the quasi-Eulerian velocity $\widehat{\mathbf u}$ is,
\begin{equation}
   \overline{D}^L \hu_i + \epsilon_{i 3 j} f_3
   \overline{u}^L_j
+\frac{\partial }{\partial x_i}\left(\frac{\overline{p}^L}{\rho_w}
-\frac{\overline{u^l_j u^l_j}}{2}\right)-\widehat{X}_i + g
\delta_{i3} =  P_j \frac{\partial \overline{u}^L_j}{\partial x_i},
\label{GLM_Gro}
\end{equation}
where the Lagrangian derivative $D^L$ is a derivative following the fluid at
the Lagrangian mean velocity $\overline{u}^L$, $p$ is the full dynamic
pressure, $\delta$ is Kronecker's symbol, and the viscous and/or turbulent
force $\widehat{\mathbf X}$ is
 defined by
\begin{equation}
 \widehat{X}_i=\overline{X}^L_i+\overline{\frac{\partial \xi_j}{\partial
 x_i}\left(\overline{X}_j^L-X_j\right)}.\label{Turb_closure}
\end{equation}

These exact equations will now be approximated using (\ref{p})-(\ref{xi3}). We
first evaluate the wave forcing terms in (\ref{GLM_Gro}) using monochromatic
waves, with a surface elevation variance $E=a^2/2$. The result for random waves
follows by summation over the spectrum and replacing $E$ with the spectral
density $E(\kb)$.

We first consider the vertical momentum balance, giving the
pressure field. It should be noted that the Lagrangian mean
Bernoulli head term $u^l_j u^l_j/2$ differs from its Eulerian
counterpart $u'_j u'_j/2$ by a term $K_2$, which arises from the
correlation of the mean current perturbation at the displaced
position $\xb+\xi$, with the wave-induced velocity, i.e. the
second term in (\ref{ulalpha}). Eqs. (\ref{p})--(\ref{xi3}) give
\begin{equation}
   \frac{1}{2}\left(u^l_j
u^l_j\right) = \frac{g k E}{2} \left[F_{CC} F_{CS} + F_{SC} F_{SS}
\right]+K_2,
\end{equation}
with
\begin{equation}
 K_2=\widetilde{u}_\alpha \xi_3 \frac{\partial \overline{u}_\alpha}{\partial z}
 + \frac{\xi_3^2}{2} \left|\frac{\partial \overline{\ub}}{\partial z}\right|^2= E \frac{\sigma}{k}  \kb \bcdot \frac{\partial
\widehat{\ub}}{\partial z} m F_{CS}F_{SS} +\frac{ E}{2}\left|\frac{\partial
\overline{\ub}}{\partial z}\right|^2 m^2 F^2_{SS} \label{K2p}.
\end{equation}


The vertical momentum equation (\ref{GLM_Gro}) for $\hw=\hu_3$ is,
\begin{eqnarray}\label{wL1}
   \frac{\partial \hw }{\partial t} &+ &\hw \frac{\partial \hw}{\partial z}
   +  P_3 \frac{\partial \hw}{\partial z} + \left(\hu_\beta + P_{\beta}\right) \frac{\partial \hw}{\partial x_\beta}
     +\frac{1}{\rho_w} \frac{\partial \overline{p}^L}{\partial z} + g \nonumber\\
     & =& \frac{\partial }{\partial z} \left[\left(\overline{{\widetilde u}_\alpha {\widetilde u}_\alpha}+\overline{\widetilde{w}^2}\right)/2+K_2\right]
     + P_{\beta} \frac{\partial }{\partial z}\left(\hu_\beta+P_{\beta} \right) + P_3 \frac{\partial }{\partial z}\left(\hu_3+P_3 \right),\label{GLM_vert}
\end{eqnarray}
For small bottom slopes we may neglect the last term, but we
rewrite it in order to compare with other sets of equations. Now
using the lowest order wave solution (\ref{u1})--(\ref{xi3}), eq.
(\ref{GLM_vert}) transforms to
\begin{eqnarray}
     \frac{1}{\rho_w} \frac{\partial }{\partial z} \left[\overline{p}^L+ \rho_w g z
     - \rho_w \frac{\sigma^2  E}{2} \left(F_{CS}^2 +F_{SS}^2\right) -\rho_w K_2 \right]& =&
    -\frac{\partial \hw}{\partial t} - \hw \frac{\partial \hw}{\partial z}\nonumber \\
   - \left(\hu_\beta  + P_\beta \right) \frac{\partial \widehat{w}}{\partial x_\beta}
   +  P_{\beta} \frac{\partial }{\partial z} \left(\hu_\beta+P_{\beta} \right)& + & P_3 \frac{\partial }{\partial z} \left(\hw+P_3 \right).\label{wL2}
\end{eqnarray}
We add to both sides the depth-uniform term $- \sigma^2 E
\left(F_{CC}^2  - F_{SS}^2\right)/2$, and integrate over $z$ to
obtain
\begin{equation}
    \frac{\overline{p(z)}^L}{\rho_w}=- g \left[ \left(z-z_s\right) - k  E F_{CC} F_{CS}
    \right]+K_2+K_1-\frac{g k   E}{4\sinh(2kD)}
    \label{pbarLsurf}
\end{equation}
where the hydrostatic hypothesis (H6, see above) has be made for
the mean flow. The depth-integrated vertical component of the
vortex-like force $K_1$ is defined by
\begin{equation}
K_1=-\int_z^{\zL} P_{\beta} \frac{\partial }{\partial
    z'} \left(\hu_\beta+P_{\beta} \right) \dr z'+\int_z^{\zL} P_{3} \frac{\partial }{\partial
    x_\beta}  \left(P_{\beta} \right) \dr z', \label{K1}
\end{equation}
where eq. (\ref{P3}) has been used. The integration constant $z_s$
is given by the surface boundary condition
\begin{equation}
    \overline{p(\zeta)}^L=-\rho_w g \left(\zL -z_s -k E  F_{CC} F_{CS} - K_2(\overline{\zeta}^L)/g \right) =
    \overline{p}_a.
\end{equation}
Using (\ref{zetaL}) we find that $z_s= \overline{\zeta} +\overline{p}_a/(\rho_w
g)- K_2(\overline{(\zeta)}^L)/g$ and (\ref{pbarLsurf}) becomes
\begin{equation}
    \frac{\overline{p}^L}{\rho_w}=\frac{\overline{p}^H}{\rho_w}+ g k E F_{CC}
F_{CS}+K_1+K_2- K_2(\overline{\zeta}^L), \label{pbar2L}
\end{equation}
with $p^H$ the hydrostatic pressure defined equal to the mean atmospheric
pressure at the mean sea surface, $p^H= \rho_w g (\zb-z)+\overline{p}_a$.

Below the wave troughs the Stokes correction for the pressure (\ref{Stokes
correction}) gives the Eulerian-mean pressure
\begin{equation}
    \overline{p}=\overline{p}^L -
\rho_w g k m E \left( F_{CS} F_{CC} + F_{SS} F_{SC}+ \frac{\kb}{ k
\sigma}\bcdot \frac{\partial \overline{\ub}}{\partial z} m F_{SS} F_{CC}
\right). \label{pStokes}
\end{equation}
Thus equation (\ref{pbarLsurf}) gives the following relationship, valid to
order $\varepsilon_1^2$ below the wave troughs, between the Eulerian-mean
pressure $\overline{p}$ and $\overline{p}^L$,
\begin{eqnarray}
 \overline{p} &=& p^H - \rho_w g k E
 F_{SS} F_{SC} + \rho_w \left( K_1-K_2(\overline{\zeta}^L)+ \frac{
E}{2}\left|\frac{\partial \overline{\ub}}{\partial z}\right|^2 m^2
F^2_{SS} \right)\nonumber \\
& & + \rho_w  g k (1-m) E F_{CC} F_{CS}.\label{pbar}
\end{eqnarray}

For a spectrum of random waves, the modified pressure term that
enters the horizontal momentum equation may be written as
\begin{equation}
   \widehat{p} \equiv \overline{p}^L-\frac{\rho_w \overline{u^l_j u^l_j}}{2}- P_j\frac{\partial \uL_i}{\partial z}=p^H+\rho_w S^{\mathrm{J}}+\rho_w S^{\mathrm{shear}}, \label{phat}
\end{equation}
with the depth-uniform wave-induced kinematic pressure term
\begin{equation}
S^{\mathrm{J}}= g \int_{\mathbf{k}} \frac{k E({\mathbf k})}{\sinh
2kD}
 {\mathrm
d}{\mathbf{k}}  \label{SJ3D}
\end{equation}
and a shear-induced pressure term, due to the integral of the vertical
component of the vortex force  $K_1$, and $K_2(\overline{\zeta}^L)$,
\begin{eqnarray}
S^{\mathrm{shear}}& = &-\int_\kb  E({\mathbf k}) \left( \frac{\sigma}{k}
k_\beta m \frac{\partial \hu_\beta(\overline{\zeta}^L)}{\partial z} \tanh(k D)
 +\frac{m^2}{2}\left|\frac{\partial
\widehat{\ub}}{\partial z}(\overline{\zeta}^L)\right|^2  \right) \dr \kb  \nonumber\\
 &+&  \int_\kb \int_z^{\zL}\left[ P_3(\kb) \frac{\partial P_{\beta}(z',\kb)}{\partial x_\beta}
    -P_{\beta}(\kb)
\frac{\partial \left[\hu_\beta(z')+ P_{\beta}(\kb)\right]
}{\partial z'} \right] \dr z' \dr \kb . \label{Sshear}
\end{eqnarray}

Now considering the horizontal momentum equations, we rewrite (\ref{GLM_Gro})
for the horizontal velocity,
\begin{eqnarray}
   \frac{\partial \widehat{u}_\alpha}{\partial t} &+& \left( \widehat{u}_\beta + P_{\beta}\right)
   \frac{\partial \widehat{u}_\alpha}{\partial x_\beta}
   +\widehat{w} \frac{\partial \widehat{u}_\alpha}{\partial z} +\epsilon_{\alpha 3 \beta} f_3
    \left( \widehat{u}_\beta + P_{\beta}\right)
+\frac{1}{\rho_w}\frac{\partial p^H}{\partial x_\alpha} \nonumber\\
 &=&
-\frac{\partial }{\partial
x_\alpha}\left(S^{\mathrm{J}}+S^{\mathrm{shear}}\right) +
P_{\beta} \frac{\partial \widehat{u}_\beta}{\partial x_\alpha}
-P_3 \frac{\partial \widehat{u}_\alpha}{\partial z} +
\widehat{X}_\alpha \label{GLM_Gro2},
\end{eqnarray}


 Grouping all $ P_{\beta}$ terms, as in Garrett
(1976\nocite{Garrett1976} eq. 3.10 and 3.11), leads to an expression with the
`vortex force' $\epsilon_{\alpha 3
   \beta} \omega_3 P_{\beta}$. This force is the
   vector product of the wave pseudo-momentum $\Pb$ and mean flow
vertical vorticity $\omega_3$. Equation (\ref{GLM_Gro2}) transforms to
\begin{eqnarray}
   \frac{\partial \widehat{u}_\alpha}{\partial t} + \widehat{u}_\beta
   \frac{\partial \widehat{u}_\alpha}{\partial x_\beta}
   + \widehat{w} \frac{\partial \widehat{u}_\alpha}{\partial z}
   &+& \epsilon_{\alpha 3
   \beta}\left[f_3 \widehat{u}_\beta+ \left(  f_3 + \omega_3\right) P_{\beta}\right]
+\frac{1}{\rho_w}\frac{\partial p^H}{\partial x_\alpha}\nonumber \\
&=& -\frac{\partial }{\partial x_\alpha}
\left(S^{\mathrm{J}}+S^{\mathrm{shear}} \right)- P_3
\frac{\partial \widehat{u}_\alpha}{\partial z}+ \widehat{X}_\alpha
\label{GLM_Gro3}.
\end{eqnarray}
The vortex force is a momentum flux divergence that compensates for the change
in wave momentum flux due to wave refraction over varying currents, and
includes the flux of momentum resulting from $\widehat{\ub}$ momentum advected
by the wave motion (Garrett 1976\nocite{Garrett1976}).

The turbulent closure is the topic of ongoing research and will not be
explicitly detailed here. We only note that it differs in principle from the
closure of the aGLM equations of Groeneweg (1999\nocite{Groeneweg1999}), which
could be extended to include the second term in eq. (\ref{Turb_closure}).
A proper closure involves a full discussion of the distortion of turbulence by
the waves when the turbulent mixing time scale is larger than the wave period
(e.g. Walmsley and Taylor 1996, Janssen
2004\nocite{Walmsley&Taylor1996,Janssen2004}, Teixeira and Belcher
2002\nocite{Teixeira&Belcher2002}). One should consider with caution the rather
bold but practical assumptions of Groeneweg (1999) who used a standard
turbulence closure to define the viscosity that acts upon the wave-induced
velocities, or the assumption of Huang and Mei (2003\nocite{Huang&Mei2003}) who
assumed that the eddy viscosity instantaneously adjusts to the passage of
waves. These effects may have consequences on the magnitude of wave attenuation
through its interaction with turbulence, and the resulting vertical profile of
$\widehat{X}_\alpha$. Here we only note that any momentum lost by the wave
field should be gained by either the atmosphere, the bottom or the mean flow.
Thus a possible parameterization for the diabatic source of momentum is
\begin{equation}
\widehat{X}_\alpha = \frac{\partial R_{\alpha \beta}}{\partial x_\beta} +
\frac{\partial }{\partial z}\left(K_z \frac{\partial
\widehat{u}_\alpha}{\partial z}\right) - T^{\mathrm{wc}}_\alpha -
T_\alpha^{\mathrm{turb}} - T^{\mathrm{bfric}}_\alpha,\label{closure_param}
\end{equation}
with $R_{\alpha \beta}$ the horizontal Reynolds stress, and $K_z$ a vertical
eddy viscosity, while the last three terms correspond to the dissipative
momentum flux from waves to the mean flow, through whitecapping,
wave-turbulence interactions, and bottom friction. Although the momentum lost
by the waves via bottom friction was shown to eventually end up in the bottom
(Longuet-Higgins 2005\nocite{Longuet-Higgins2005}), the intermediate
acceleration of the mean flow, also known as Eulerian streaming, is important
for sediment transport, and should be included with a vertical profile of
$T^{\mathrm{bfric}}_\alpha$ concentrated near the bottom, provided that the
wave boundary layer is actually resolved in the 3D model (e.g. Walstra et al.
2001).

The GLM mass conservation writes
\begin{equation}
   \frac{\partial \left(J\right)}{\partial t}+
\frac{\partial \left( J \overline{u}^L_\alpha \right)}{\partial x_\alpha}
+\frac{\partial \left( J \overline{w}^L \right)}{\partial z} =
0,\label{GLMmass}
\end{equation}
where the Jacobian $J$ is the determinant of the coordinate transform matrix
${\left(\delta_{ij}+\partial \xi_i/\partial x_j\right)}$  from Cartesian
coordinates to GLM.
 (Andrews and McIntyre 1978a, eq. (4.2)-(4.4) with $\rho^\xi=\rho_w$).

\subsection{\textit{glm}2-RANS equations in $z$-coordinates}
Equations (\ref{GLM_Gro3}) and (\ref{GLMmass}) hold from $z=-h$ to
$z=\zL$, which covers the entire `GLM water column'. All terms in
(\ref{GLM_Gro3}) are defined as GLM averages, except for the
hydrostatic pressure $p^H$ which does correspond to the Eulerian
mean position.

For practical numerical modelling, it is however preferable that
the height of the water column does not change with the local wave
height. We will thus transform eq. (\ref{GLM_Gro3}), except for
$p^H$, by correcting for the GLM-induced vertical displacements.
This will naturally remove the divergence of the GLM flow related
to $J\neq 1$. The GLM vertical displacement $\overline{\xi}_3^L$
is a generalization of eq. (\ref{zetaL})
\begin{eqnarray}
   \overline{\xi}_3^L (x,z,t)= \int_{\mathbf k} E(\kb)
m\left[k \frac{\sinh\left[2k(z+h)\right]}{2 \sinh^2(k D)}+m
\frac{\sinh^2\left[k(z+h)\right]}{\sinh^2(k D)} \frac{\kb}{\sigma} \bcdot
\frac{\partial \overline{u}_\alpha}{\partial z} \right]{\mathrm d}\kb.\nonumber
\\\label{s2Gdef}
\end{eqnarray}
and the Jacobian is $J=1+J_2+O(\varepsilon_1^3)$. Because the GLM does not
induce horizontal distortions, a vertical distance ${\mathrm d}z^\prime=J
{\mathrm d}z$ in GLM corresponds to a Cartesian distance ${\mathrm d}z$,
giving,
\begin{eqnarray}
J_2 & = & -\frac{\partial \overline{\xi}_3^L}{\partial z}.
\end{eqnarray}
One may note that
\begin{equation}
   \int_{-h}^{\zL} J {\mathrm d}z =  \overline{\zeta}^L + h -
\overline{\xi}_3^L(0) = D.
\end{equation}

We now implicitly define the vertical coordinate $z^\star$ with
\begin{equation}
   s=z^\star + \overline{\xi}_3^L \label{sdef}
\end{equation}
Any field $\phi(x_1,x_2,z,t)$ transforms to
$\phi^\star(x_1^\star,x_2^\star,z^\star,t^\star)$ with
\begin{eqnarray}
\frac{\partial \phi}{\partial t} &=& \frac{\partial \phi^\star}{\partial
t^\star}
- \frac{s_t}{s_z}\frac{\partial \phi^\star}{\partial z^\star} \label{sigma_t}\\
\frac{\partial \phi}{\partial x_\alpha} &=& \frac{\partial \phi^\star}{\partial
x_\alpha^\star}
- \frac{s_\alpha}{s_z}\frac{\partial \phi^\star}{\partial z^\star}\label{sigma_alpha}\\
\frac{\partial \phi}{\partial z} &=& \frac{1}{s_z}\frac{\partial
\phi^\star}{\partial z^\star}\label{sigma_z}
\end{eqnarray}
with $s_t$, $s_z$ and $s_\alpha$ the partial derivatives of $s$ with respect to
$t^\star$, $z^\star$ and $x_\alpha^\star$, respectively. The coordinate
transform was built to obtain the following identity
\begin{equation}
   s_z J = 1+O\left(\varepsilon_1^3\right).\label{s2J}
\end{equation}

Removing the $\star$ superscripts from now on, the mass conservation
(\ref{GLMmass}) multiplied by $s_z$ may be written as
\begin{equation}
  \frac{\partial \left( \uL_\alpha\right)}{\partial x_\alpha}
+\frac{\partial \left(W\right)}{\partial z} = 0, \label{GLMmassz}
\end{equation}
where the  vertical  velocity,
\begin{equation}
   W =J  \left[ \overline{w}^L -  \overline{u}^L_\alpha s_{\alpha} - s_t \right]
   = \widehat{w}\frac{1+O(\varepsilon)}{\partial \overline{\xi}_3^L/\partial z},
\label{DefOmega}
\end{equation}
is the Lagrangian mass flux through horizontal planes.

Neglecting terms of order $\varepsilon_1^3$ and higher, the
product of (\ref{GLM_Gro3}) and $s_z J$ is re-written as,
\begin{eqnarray}
  \frac{\partial \widehat{u}_\alpha}{\partial t} & + & \widehat{u}_\beta \frac{\partial
   \widehat{u}_\alpha}{\partial x_\beta}
   + \widehat{w} \frac{\partial \widehat{u}_\alpha }{\partial
   z}+ \epsilon_{\alpha 3
   \beta}\left[ f_3 \widehat{u}_\beta + \left(  f_3 + \omega_3\right) P_{\beta}\right]
   + \frac{\partial p^H}{\partial x_\alpha} \nonumber \\
    &=&  -\frac{\partial }{\partial x_\alpha}\left(S^{\mathrm{J}}+S^{\mathrm{shear}} \right) - P_3
\frac{\partial \widehat{u}_\alpha}{\partial z} +
\widehat{X}_\alpha \label{GLM2z},
\end{eqnarray}
with
\begin{eqnarray}
   \widehat{w}  &=& J\left[\overline{w}^L - \widehat{u}_\alpha s_{\alpha} - s_{t}
   \right]- P_3
   = W-P_3+ J P_{\alpha} s_{\alpha} \nonumber \\
   &=& W -P_3+ O(\sigma \varepsilon_1^4 \varepsilon_2/k),
\label{DefOmegaEul}
\end{eqnarray}
the quasi-Eulerian advection velocity through horizontal planes.
From now on we shall use exclusively these $glm2z$-RANS equations
in $z$ coordinate, with a non-divergent GLM velocity field
$\overline{\ub}^L$.

Using eq. (\ref{P3}), we may re-write (\ref{GLMmassz})  as
\begin{equation}
  \frac{\partial  \widehat{u}_\alpha}{\partial x_\alpha}
+\frac{\partial \hw }{\partial z} = 0. \label{GLMmasszb}
\end{equation}

\subsubsection{Surface boundary conditions}
Taking an impermeable boundary, the kinematic boundary condition  is given by
Andrews and McIntyre (1978a, section 4.2),
\begin{equation}
   \frac{\partial \zL}{\partial t} + \overline{u}^L_\alpha \frac{\partial \zL}{\partial x_\alpha}
= \overline{w}^L \quad \mathrm{at} \quad z = \zL. \label{GLMsfcbcO}
\end{equation}
It is transformed to $z$ coordinates as
\begin{equation}
   \frac{\partial \zb}{\partial t} + \overline{u}^L_\alpha \frac{\partial \zb}{\partial x_\alpha}
= W = \hw + P_3 \quad \mathrm{at} \quad z = \zb. \label{GLMsfcbc}
\end{equation}

When the presence of air is considered, it should be noted that the GLM
position is discontinuous in the absence of viscosity, because the Stokes
corrections for $\zeta$ have opposite signs in the air and in the water. This
discontinuity arises from the discontinuity of the horizontal displacement
$\xi_\alpha$ (air and water wave-induced motions are out of phase). A proper
treatment would therefore require to resolve the viscous boundary layer at the
free surface. This question is left for further investigation. However, we note
that due to the large wind velocities and possibly large surface currents
unrelated to wave motions, a good approximation is given by neglecting the
Stokes corrections for the horizontal air momentum,
\begin{equation}\widehat{u}^+_\alpha = \widehat{u}^-_\alpha +
 P^-_{\alpha},
\end{equation}
where the $-$ and $+$ exponents refer to the limits when approaching the
boundary from below and above, respectively.

For the mean horizontal stress, we use the results of Xu and Bowen
(1994\nocite{xu&Bowen1994}),
\begin{equation}
\tau_{\alpha} = \overline{S_{nn} n_\alpha}  + \overline{S_{ns} n_3}  \quad
\textrm{at} \quad z=\zb \label{taucontsurf}
\end{equation}
with $\mathbf{S}$ the stress tensor, with normal $S_{nn}$ and shear $S_{ns}$
stresses on the surface, generally defined by
\begin{equation} S_{ij} = -p \delta _{ij} + \rho_w \nu \left(\frac{\partial u_i}{\partial x_j} +
\frac{\partial u_j}{\partial x_i}\right) \label{viscstress},
\end{equation}
with $\nu$ the kinematic viscosity, and the local unit vector normal to the
surface, to first order in $\varepsilon_1$,
\begin{equation}
{\mathbf n} = (0,0,1) - \left( \frac{\partial \zeta}{\partial
x_1},\frac{\partial \zeta}{\partial x_2},0\right)\label{n}.
\end{equation}

Taking the Lagrangian mean of (\ref{taucontsurf}), one obtains,
\begin{equation}
\tau^a_\alpha = \overline{\tau_{\alpha}}^L =\tau^w_{\alpha} + \rho_w \nu
\frac{\partial \widehat{u}_\alpha}{\partial z} + \rho_w \nu \frac{\partial
P_{\alpha}}{\partial z} \quad \textrm{at} \quad z = \zb, \label{surfstress1b}
\end{equation}
where $\tau^a_{\alpha}$ is the total air-sea momentum flux (the wind stress),
as can be measured above the wave-perturbed layer (e.g. Drennan et al.
1999\nocite{Drennan&al.1999b}). $\tau^w_{\alpha}$ is the $\alpha$ component of
the wave-supported stress due to surface-slope pressure correlations,
\begin{equation}
\tau^w_{\alpha}=\overline{p \frac{\partial \zeta}{\partial x_\alpha}}^L.
\label{tauw}
\end{equation}

The second viscous term $\rho_w \nu {\partial
P_{\alpha}}/{\partial z}$ was estimated using the GLM average of
wave orbital shears (Ardhuin and Jenkins 2006), it is the
well-known virtual wave stress (e.g. Xu and Bowen 1994, eq. 18).
That stress corresponds to wave momentum lost due to viscous
dissipation, and it can be absorbed into the boundary conditions
because it is concentrated within a few
 millimeters from the surface (Banner et Peirson 1998\nocite{Banner&Peirson1998}).  At the base of the viscous layer of thickness
$\delta_s$, (\ref{surfstress1b}) yields, using an eddy viscosity $K_z$,
\begin{equation}
\tau^a_{\alpha}-\tau^w_{\alpha}-\rho_w \nu \frac{\partial P_{\alpha}}{\partial
z} = \rho_w K_z \frac{\partial \widehat{u}_\alpha}{\partial z}  \quad
\textrm{at} \quad z = -\delta_s. \label{surfstress2b}
\end{equation}

\subsubsection{Bottom boundary conditions}
The same approach applies to the bottom boundary conditions. The kinematic
boundary condition writes
\begin{equation}
   \frac{\partial \overline{h}^L}{\partial t} + \left(\hu_\alpha + P_\alpha \right) \frac{\partial \overline{h}^L}{\partial x_\alpha}
=  \left(\hw + P_3 \right) \quad \mathrm{at} \quad z =
-\overline{h}^L. \label{GLMbotbc}
\end{equation}
If an adherence condition is specified at the bottom, which shall be used
below, the bottom boundary condition further simplifies as $\overline{h}^L=h$.
It may also simplify under the condition that the wave amplitude is not
correlated with the small scale variations of $h$, which is not generally the
case (e.g. Ardhuin and Magne 2007\nocite{Ardhuin&Magne2007}). For the dynamic
boundary conditions, pressure-slope correlations give rise to a partial
reflection of waves, that may be represented by a scattering stress (e.g. Hara
and Mei 1987\nocite{Hara&Mei1987}, Ardhuin and Magne 2007). This stress
modifies the wave pseudo-momentum without any change of wave action (see also
Ardhuin 2006\nocite{Ardhuin2006b}).

The effect of bottom friction is of considerable interest for sediment dynamics
and deserves special attention. For the sake of simplicity, we shall here use
the conduction solution of Longuet-Higgins for a constant viscosity over a flat
sea bed as given in the appendix to the proceedings of Russel and Osorio
(1958\nocite{Russel&Osorio1958}). We shall briefly consider waves propagating
along the $x$-axis, and we assume that the mean current in the wave bottom
boundary layer (WBBL)  is at most of the order of the wave orbital velocity
outside of the WBBL. Instead of (\ref{u1})--(\ref{xi3}) the orbital wave
velocity and displacements near the bottom take the form,
\begin{eqnarray}
{u_1}& =& u_0 \left[\cos \psi-\er^{-\widehat{z}} \cos(\psi-\widehat{z})\right]  \label{ubl}\\
{w}& =& \frac{u_0 k \delta_f}{2} \left[2 \widehat{z} \sin\psi  - \sin
(\psi-\widehat{z})\er^{-\widehat{z}} + \sin \psi + \cos
(\psi-\widehat{z})\er^{-\widehat{z}}
- \cos \psi \right]\\
{\xi_1}& =& -\frac{u_0}{\omega}   \left[\sin\psi -  \sin(\psi-\widehat{z})\er^{-\widehat{z}}\right] \\
{\xi_3}& =& \frac{u_0 k \delta_f}{2 \omega}   \left[2 \widehat{z} \cos \psi -
 \cos(\psi-\widehat{z}) \er^{-\widehat{z}} + \cos
\psi+\sin(\psi-\widehat{z}) \er^{-\widehat{z}}- \sin \psi \right] \label{xi3bl}
\end{eqnarray}
where $\psi=k x -\omega t$ is the wave phase,
$\delta_f=\left(2\nu/\omega\right)^{1/2}$ is the depth scale for the boundary
layer, $\widehat{z}=(z+h)/\delta_f$ is a non-dimensional vertical coordinate,
$u_0 = a \sigma/\sinh(kD)$ is the orbital velocity amplitude outside the
boundary layer.

 Based on these velocities and displacements, the wave pseudo-momentum $P$, is
\begin{equation}
P_1=\overline{-\xi_{1,1} u_1 -\xi_{3,1} w}= \frac{u_0^2}{2 C}
\left[1+\er^{-2\widehat{z}} \cos(2 \widehat{z})-2\cos \widehat{z}
\er^{-\widehat{z}}\right].
\end{equation}
This is equal to the Stokes drift $\overline{u}^S=\overline{u_{1,1} \xi_1
+u_{1,3} \xi_3}$ computed by Longuet-Higgins. Besides, the rate of wave energy
dissipation induced by bottom friction is $S_{\mathrm{bfric}}=\rho_w \omega
u_0^2/2$ giving a bottom friction stress
$\int_{-h}^{\infty}T^{\mathrm{bfric}}_\alpha \dr z = k_\alpha
S_{\mathrm{bfric}}/(\rho_w \sigma)$.

Generalizing this approach to a turbulent bottom boundary layer (e.g.
Longuet-Higgins 2005) one may replace the constant viscosity with a
depth-varying eddy viscosity. If the wave bottom boundary layer (WBBL) is
resolved, $\overline{\tau}^b_{\alpha}$ will also include the momentum lost by
waves through bottom friction, as given by the depth-integral of
$T^{\mathrm{bfric}}_\alpha$. One may estimate $P$ from the vertical profiles of
the  wave orbital velocities  $\widetilde{u}_\alpha$ and $\widetilde{w}$, and
the modified pressure (\ref{phat}) has to be corrected for the change in wave
orbital velocities in the WBBL. Many WBBL models are available for estimating
these wave-induced quantities.

If the bottom boundary layer is not resolved, on may take the
lowest model level at the top of the wave boundary layer. The
bottom stress may then be computed from a parameterization of the
bottom roughness $z_{0 a^\prime}$ (e.g. Mathisen and Madsen 1996,
1999\nocite{Mathisen&Madsen1996a,Mathisen&Madsen1999}), which
relates the bottom stress
\begin{equation}
\overline{\tau}^b_{\alpha} = - \rho_w u_{\star c}^2
\frac{\widehat{u}_\alpha}{\widehat{u}},
\end{equation}
to the current velocity $\widehat{u}_\alpha$ at the lowest model level $z$,
\begin{equation} \widehat{u}_\alpha = \kappa u_{\star
c} \ln \left[\frac{z+h}{z_{0 a^\prime}}\right],  \quad \textrm{for} \quad z+h<
\delta_f \label{BBL1b}.
\end{equation}
Then the near-bottom velocity $\widehat{u}_\alpha$ should be taken
equal to the Eulerian streaming velocity $\sim 1.5 P_\alpha$ (see
e.g. Marin 2004\nocite{Marin2004}, for turbulent cases with
rippled beds). Further, in this case the bottom stress
$\overline{\tau}^b_{\alpha}$ should not include the depth integral
of $T^{\mathrm{bfric}}_\alpha$. This latter remark also applies to
depth-integrated equations. Indeed,
$\tau_\alpha^{wb}=\int_{-h}^{-h+\delta_f}T^{\mathrm{bfric}}_\alpha
\dr z$ is a flux of momentum into the bottom due to wave bottom
friction, $\tau_\alpha^{wb}$ does not participate in the momentum
balance that gives rise to a sea level set-down and set-up
(Longuet-Higgins 2005).

\section{Relations between the present theory and known equations}
\subsection{Depth-integrated GLM for a constant density $\rho_w$} Using
(\ref{GLMsfcbc}) the mass conservation equation in $z$ coordinates
(\ref{GLMmassz}) classically gives (e.g. Phillips 1977\nocite{Phillips1977})
\begin{equation}
   \frac{\partial}{\partial t} \int_{-h}^{\zb} \rho_w  {\mathrm d} z
= - \frac{\partial}{\partial x_\alpha}\int_{-h}^{\zb} \rho_w \uL_\alpha
{\mathrm d} z \label{GLMmass_int}
\end{equation}
which is exactly the classic shallow-water mass conservation for constant
density,
\begin{equation}
  \frac{\partial D}{\partial t}
= - \frac{\partial M_\alpha}{\partial x_\alpha} \label{mass_int},
\end{equation}
with the depth-integrated volume flux vector\footnote{Phillips (1977) uses the
notation $\widetilde{\mathbf M}$ instead of $\mathbf M$, and $\mathbf M$
instead of $\mathbf M^w$.} $\mathbf M$ defined by
\begin{equation}
{\mathbf M}=\int_{-h}^{\overline{\zeta}} \overline{{\mathbf u}}^L {\mathrm d}
z.\label{Mtot}
\end{equation}

In the momentum equation, the advection terms may be transformed
in flux form using mass conservation. However, because some of the
original GLM advection terms are included in the vortex force, the
remaining terms do not simplify completely. Using
(\ref{GLMmasszb}) one has,
\begin{eqnarray}
    &\rho_w &\left[\frac{\partial \hu_\alpha}{\partial t} +
\hu_\beta \frac{\partial \hu_\alpha}{\partial x_\beta} + \hw
\frac{\partial \hu_\alpha}{\partial z}\right]+ P_3
\frac{\partial \hu_\alpha}{\partial z} \nonumber \\
 & =&\frac{\partial }{\partial t}\left(\rho_w \hu_\alpha\right) +
\frac{\partial }{\partial x_\beta}\left(\rho_w  \hu_\beta
\hu_\alpha \right)  +\frac{\partial }{\partial z} \left[\rho_w
\left(\hw +P_3\right)\hu_\alpha \right] - \hu_\alpha
\frac{\partial P_3}{\partial z} . \label{GLMadv_flux}
\end{eqnarray}
Using (\ref{GLMsfcbc}), (\ref{GLMbotbc}) and (\ref{P3}), and after
integration by parts, these advection terms integrate to
\begin{equation}
    \frac{\partial M^m_\alpha}{\partial t} + \frac{\partial}{\partial x_\beta}\left(\int_{-h}^{\zb}\rho_w \hu_\alpha \hu_\beta {\mathrm d}
    z\right)
 + u_{A \alpha} \frac{\partial M^w_{\beta}}{\partial x_\beta}  +
\frac{\partial u_{A \alpha}}{\partial x_\beta}M^w_{\beta} -
\int_{-h}^{\overline{\zeta}} P_\beta \frac{\partial
\hu_\alpha}{\partial x_\beta}{\mathrm d} z \label{GLMadv_int},
\end{equation}
where the zeroth order wave advection velocity ${\mathbf u}_{A}$
is defined by,
\begin{equation}
u_{A \alpha} M^w_{\beta} \equiv \int_{-h}^{\zb} \hu_\alpha
P_{\beta} {\mathrm d} z \label{uA},
\end{equation}
which is equal, at lowest order, to the second term in (\ref{dispersion}). The
wave-induced mass transport is the depth-integrated pseudo-momentum,
\begin{equation}
\Mb^w=\int_{-h}^{\overline{\zeta}} \Pb {\mathrm d} z.\label{Mw}
\end{equation}
Finally, the quasi-Eulerian volume flux is defined by $\Mb^m=\Mb-\Mb^w$.

For terms uniform over the depth ($\partial p^H/\partial x_\alpha$ and
$\partial S^{\mathrm{J}}/\partial x_\alpha$) the integral is simply the
integrand times the depth.

It should be noted that the depth-integrated vortex force involves
the advection velocity ${\mathbf u}_{A}$,
\begin{equation}
    \int_{-h}^{\overline{\zeta}^L} \epsilon_{\alpha 3
   \beta} \left(  f_3 + \omega_3\right) P_{\beta} {\mathrm d} z
=\epsilon_{\alpha 3
   \beta}\left(  f_3 + \Omega_3\right) M^w_{\beta}
\label{GLMvort_int},
\end{equation}
with
\begin{equation}
\Omega_3=\epsilon_{3 \alpha \beta}
 \left({\partial u_{A \beta}}/{\partial x_\alpha}-{\partial u_{A \alpha}}/{\partial
 x_\beta}\right)
\label{Omega_3}.
\end{equation}

The vertical integration of (\ref{GLM2z}) thus yields
\begin{eqnarray}
   &\frac{\partial M^m_\alpha}{\partial t} &+   \frac{\partial}{\partial x_\beta}\left(\int_{-h}^{\zb}\rho_w \hu_\alpha \hu_\beta {\mathrm d}
    z\right)
+ \epsilon_{\alpha 3 \beta} f_3  M^m_\beta
 +   D      \frac{\partial }{\partial x_\alpha} \left(\rho_w g
 \overline{\zeta}+p_a\right) \nonumber\\
 & =&  - \epsilon_{\alpha 3 \beta}   \left(f_3+\Omega_3\right)
M^w_{\beta} - u_{A \alpha} \frac{\partial M^w_{\beta}}{\partial
x_\beta} - \frac{\partial u_{A \alpha}}{\partial
x_\beta}M^w_{\beta} + \int_{-h}^{\overline{\zeta}} P_\beta
\frac{\partial
\hu_\alpha}{\partial x_\beta}{\mathrm d} z \nonumber\\
& & - D \frac{\partial S^{\mathrm{J}}}{\partial x_\alpha} -
\int_{-h}^{\zb} \frac{\partial S^{\mathrm{shear}}}{\partial
x_\alpha} \dr z - \int_{-h}^{\zb}  P_3 \frac{\partial
\widehat{u}_\alpha}{\partial z}  \dr z + X^{\rm int}.
\label{GLMint}
\end{eqnarray}
The source of momentum $X^{\rm int}$ is simply the sum of the mean
momentum fluxes at the top and bottom, and the source of momentum
due to diabatic wave-mean flow interactions (i.e. breaking and
wave-turbulence interactions).

These equations are very similar to those of Smith
(2006\nocite{Smith2006b}, eq. 2.29), our term $S^{\mathrm{J}}$ is
simply termed $J$ in Smith (2006), and $X^{\rm int}$ corresponds
to Smith's $k_i D^W$. The only differences are due to the vertical
shear in the  current. The advection velocity $u_{A \alpha}$
replaces Smith's mean flow velocity. Since $u_{A \alpha}$ is the
proper lowest order advection velocity for the wave action
(Andrews and McIntyre 1978b\nocite{Andrews&McIntyre1978b}), this
is a simple extension of Smith's result to depth-varying currents.
The term involving $S^{\mathrm{shear}}$ is also obviously absent
from Smith's equations. The last differences in (\ref{GLMint}) are
the last two terms on the second line, but they also cancel for a
depth-uniform current $\hu_\alpha$.

\subsection{Equations of McWilliams et al. (2004)}
The approach of MRL04 is in the line of perturbation theories
presented by Mei (1989\nocite{Mei1989}) for Eulerian variables and
monochromatic waves. Although the result of MRL04 corresponds to a
particular choice of the relative ordering of small parameters, it
is given to a high enough order so that it does cover most
situations at a lower order. In particular MRL04 have pushed the
expansion to order $\varepsilon_1^4$ for some terms because they
assumed a ratio $\sigma / f_3$ of order $\varepsilon_1^4$, with
$\varepsilon_1$ the wave slope. This ratio, in practice, may only
be attained for relatively steep wind waves (developed wind seas
and swells generally have slopes of the order of 0.05). They also
assumed that $\varepsilon_1^2 \sim \varepsilon_2$ (the wave
envelope varies on a scale relatively larger than the wavelength
compared to the present theory in which $\varepsilon_1 \sim
\varepsilon_2$ is possible). These authors also separated the
motion into waves, long waves and mean flow, and considered in
detail the rotational part of the wave motion caused by the
vertical shear of the current.

MRL04 thus obtained Eulerian-mean equations that only correspond
to measurable Eulerian averages under the level of the wave
troughs. Because they use an analytic continuation of the velocity
profiles across the air-sea interface, the physical interpretation
of their average is unclear between the crests and troughs of the
waves. We shall neglect here their terms of order
$\varepsilon_1^4$ (i.e. terms that involve the wave amplitude to
the power of four), which amounts to choosing a slightly different
scaling. Since we shall consider here random waves, this avoids
cumbersome considerations of the wave bispectrum.

The Eulerian-mean variables of MRL04 should be related to the
Lagrangian mean values by the Stokes corrections (\ref{Stokes
correction}), so that their horizontal Eulerian-mean velocity
$\mathbf q$ corresponds to $\overline{\mathbf
u}^L-\overline{\mathbf u}^S$. Because they have subtracted the
hydrostatic pressure with the mean water density $\rho_{w0}$,
their mean pressure $\langle p \rangle$ should be equal to the
Eulerian mean pressure $\overline{p}+\rho_{w0}gz$, with
$\overline{p}$ related to the GLM pressure via eq. (\ref{pbar}).

Absorbing the long waves in the mean flow (i.e. allowing the mean
flow to vary on a the wave group scale, see also Ardhuin et al.
2004\nocite{Ardhuin&al.2004a}), MRL04 equations for the `Eulerian'
mean velocity $(q_1,q_2,\overline{w})$ can be written as
\begin{eqnarray}
\frac{\partial q_\alpha}{\partial t}  &+ &\left(q_\beta
\frac{\partial }{\partial x_\beta} + \overline{w} \frac{\partial
}{\partial z}\right) q_\alpha + \epsilon_{\alpha 3 \beta} f_3
q_\beta + \frac{1}{\rho_w} \frac{\partial \langle p
\rangle}{\partial x_\alpha} =  -\frac{\partial}{\partial
x_\alpha}\left(\mathcal{K}_1 +\mathcal{K}_2\right) + J_\alpha
\nonumber
\label{McWill1} \\ \\
\frac{\partial \langle p \rangle}{\partial z} & = &(\rho_w-\rho_{w0}) g  - \frac{\partial }{\partial z} \left(\mathcal{K}_1 +\mathcal{K}_2\right)+ K \label{McWill2}\\
& &\frac{\partial q_\beta}{\partial x_\beta} + \frac{\partial
\overline{w}}{\partial z} =
 0 \\
\langle p \rangle& =&\rho_w g \left(\overline{\zeta}-k E F_{SC} F_{SS}\right) -
\mathcal{P}_0 \quad \mathrm{at} \quad z= 0 \label{McWill_surfp}
\\
\overline{w}& =& - w^{St} \quad \mathrm{at} \quad z= 0
\label{McWill_surfw}
\end{eqnarray}
with
\begin{eqnarray}
\mathcal{K}_1 & = & \frac{\overline{\widetilde{u}_j \widetilde{u}_j}}{2}  =
-\frac{1}{2} \left[F_{CC} F_{CS} + F_{SS} F_{SC}\right] g k E \\
J_\alpha  & = & - \epsilon_{\alpha 3 \beta} \left(  f_3 + \omega_3\right)  \overline{u}_{\beta}^S  - w^{St} \frac{\partial q_\alpha}{\partial z} \\
K & =& \overline{u}_{\beta}^S \frac{\partial q_\beta}{\partial z} \\
\mathcal{K}_2 & = & \frac{\sigma k_\beta E}{k} \int_{-h}^z \frac{\partial^2
q_\beta(z^\prime)}{\partial z^2}F_{CS}(z^\prime) F_{SS}(z^\prime)
{\mathrm d}z^\prime \\
\mathcal{P}_0 &= &O(\frac{g}{k} \varepsilon_1^4)
\end{eqnarray}
The original notations of MRL04 (see also Lane et al.
2007\nocite{Lane&al.2007}) have been translated to the notations
used above and order  $\varepsilon_1^4$ terms have been neglected.

These equations are clearly analogue to the $glm2z$-RANS equations
presented here. In particular the vertical vortex force term $K$
corresponds to our $K_1$ that gets into $S^{\mathrm{shear}}$,  the
dynamically relevant kinematic pressure pressure $\langle p
\rangle + \mathcal{K}_1 +\mathcal{K}_2$ corresponds to our
pressure $\widehat{p}$ defined by (\ref{phat}), and the vertical
Stokes velocity $w^{St}$ corresponds to our $P_3$. There are only
two differences. One is between the surface boundary conditions
for these two pressures, with a difference only due to
$\mathcal{K}_2(z=0) \neq -K_2(\overline{\zeta}^L)$. Integrating by
parts to estimate $\mathcal{K}_2(z=0)$, this difference is found
to be of the order of $g k E \varepsilon_3$. Such a difference is
of the same order as extra terms that would arise when using wave
kinematics to first order in the current curvature (Kirby and Chen
1989), and properly transforming $\widehat{u}$ in $\overline{u}$.
The second difference between MRL04 and the present equations is
that the wave pseudo-momentum $\Pb$ differs from the Stokes drift
$\overline{\mathbf u}^S$ when the current shear is large, and both
generally differ from the expression for $\overline{\mathbf u}^S$
given by MRL04. Since MRL04 took the current and wave orbital
velocity to be of the same order, in that context the difference
$\Pb-\overline{\mathbf u}^S$ is of higher order and thus the two
sets of equations are consistent in their common range of
validity.

A general comparison of 2D depth-integrated equations is discussed by Lane et
al. (2006). The present work therefore brings a further verification of their
3D form of the equations, and an extension to relatively strong currents,
possibly as large as the phase velocities. As expected, the Eulerian averages
of  McWilliams et al. (2004) are identical to the quasi-Eulerian fields in GLM
 theory,
because they obey the same equations,  except for current profile
curvature effects, which were partly neglected here. The
"Eulerian" mean current of MRL04 can thus be physically
interpreted as a quasi-Eulerian average, defined as the GLM
average minus the wave pseudo-momentum. Except for a Jacobian that
introduces relative corrections of second order in the wave slope,
this averaging is identical to the procedure used by Swan et al.
(2001). Above the trough level, this average should not be
confused with a truly Eulerian average, as obtained from in-situ
measurements for example. In such measurements the Stokes drift
would be recorded in the trough-to-crest region (figure 1.a).

\section{Limitations of the approximations}
The $glm2z$-RANS equations have been obtained from the exact GLM
equations, under 6 restricting hypotheses related to the wave
slope and Ursell number (H1 and H2), the horizontal scales of
variation of the wave amplitude (H3), the current profile (H4 and
H5) and the vertical mean velocity (H6). These hypotheses
essentially allowed us to use the linear wave-induced quantities
given by eqs. (\ref{u1})--(\ref{xi1}).  In practical conditions,
these hypotheses may not be verified and the resulting
$glm2z$-RANS equations may have to be modified. Here we
investigate the importance of H3, H2  and H1, using numerical
solutions from  an accurate coupled mode model for irrotational
wave propagation over any bottom topography, and an accurate
analytical solution for incipient breaking waves, respectively.

\subsection{Bottom slope and standing waves}
In absence of dissipation and given proper lateral boundary
conditions the flow in wave shoaling over a bottom slope is
irrotational and can thus be obtained by a numerical exact
solution of Laplace's equation with bottom, surface, and lateral
boundary conditions. For waves of small amplitudes this can be
provided by a solution to this system of equations to second order
in the wave slope. Belibassakis and Athanassoulis
(2002\nocite{Belibassakis&Athanassoulis2002}) have developed a
second order version of the National Technical University of
Athens numerical model  (NTUA-nl2) to solve this problem in two
dimensions. Here we apply their model to the simple case of
monochromatic, unidirectional waves propagating along the $x$
axis, with a topography uniform along the $y$ axis. The topography
$h(x)$ varies only for $0<x<L$ and is constant $h(x)=h_1$ for
$x<0$ and $h(x)=h_2$ for $x>L$. In that case the Eulerian mean
current $\bnabla \phi_0(\xb)$ is irrotational, and uniform over
the vertical as $x$ approaches $\pm \infty$ (e.g. Belibassakis and
Athanassoulis 2002, table 1 and figure 5). We shall further
restrict our investigation to the case of a monochromatic wave
train of known radian frequency $\omega$ and incident amplitude
$a$, giving rise to reflected and transmitted wave trains of
amplitudes $R a$ and $T a$. Numerical calculations are given for a
bottom profile as given by Roseau (1976\nocite{Roseau1976}) for
which the reflection coefficient $R$ is known analytically, thus
providing a check on the quality of the numerical solution.

The bottom is defined here by $x$ and $z$ coordinates given by the
real and imaginary part of the complex parametric function of the
real variable $x'$,
\begin{equation}
Z(x')=x + \ir z = \frac{h_1 (x'-\ir \alpha_0)+(h_2-h_1) \ln (1+\er^{x'-\ir
\alpha_0})}{\alpha_0}.\label{Roseau}
\end{equation}
We choose $h_1=6$~m and $h_2=4$~m and a wave frequency of 0.19~Hz
($\omega=1.2$~rad~s$^{-1}$). For $\alpha_0=15 \pi/180$ the maximum bottom slope
is $\varepsilon_2=2.6\times 10^{-2}$ (figure 1), and the reflection coefficient
for wave amplitude is $R=1.4\times 10^{-9}$ (Roseau 1976\nocite{Roseau1976}),
so that reflected waves may be neglected in the momentum balance.
\begin{figure}[htb]
 \vspace{9pt}
\centerline{\includegraphics[width=\textwidth]{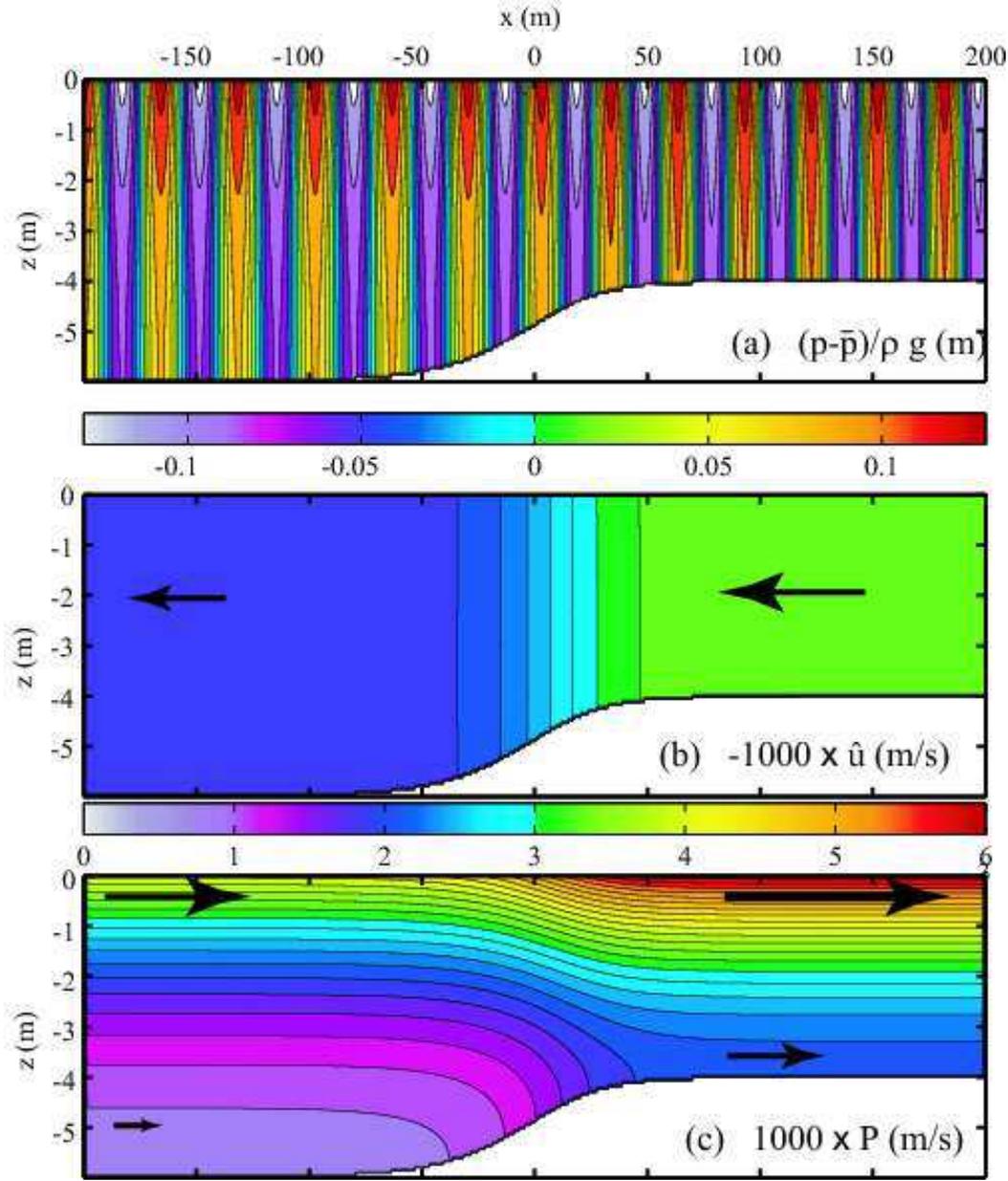}}
 \caption{(a) Instantaneous pressure perturbation $(p-\overline{p})/(\rho_w g)$ given by the NTUA-nl2 model (Belibassakis and Athanassoulis 2002),
 including the second order Stokes
 component in waves with amplitude $a=0.12$~m,
 over the bottom given by eq. (\ref{Roseau}).
 (b) Mean current $-\hu$,
 and (c) horizontal wave pseudo-momentum $P_1$ estimated from eq. (\ref{P}), and verified to be
 equal to the Stokes drift. Arrows indicate the flow directions.}
 \label{figNTUA1}
\end{figure}
Due to the shoaling of the incident waves, the mass transport
induced by the waves increases in shallow water, and thus the mean
current must change in the $x$ direction to compensate for the
divergence in the wave-induced mass transport. We shall further
take a zero-mean surface elevation as $x\rightarrow -\infty$. The
second order mean elevation is obtained as a result of the model.
We also verified that the vertical wave pseudo-momentum
compensates for the divergence of the horizontal component so that
in this case for linear waves the wave pseudo-momentum is
non-divergent (figure 3).
\begin{figure}[htb]
 \vspace{9pt}
\centerline{\includegraphics[width=\textwidth]{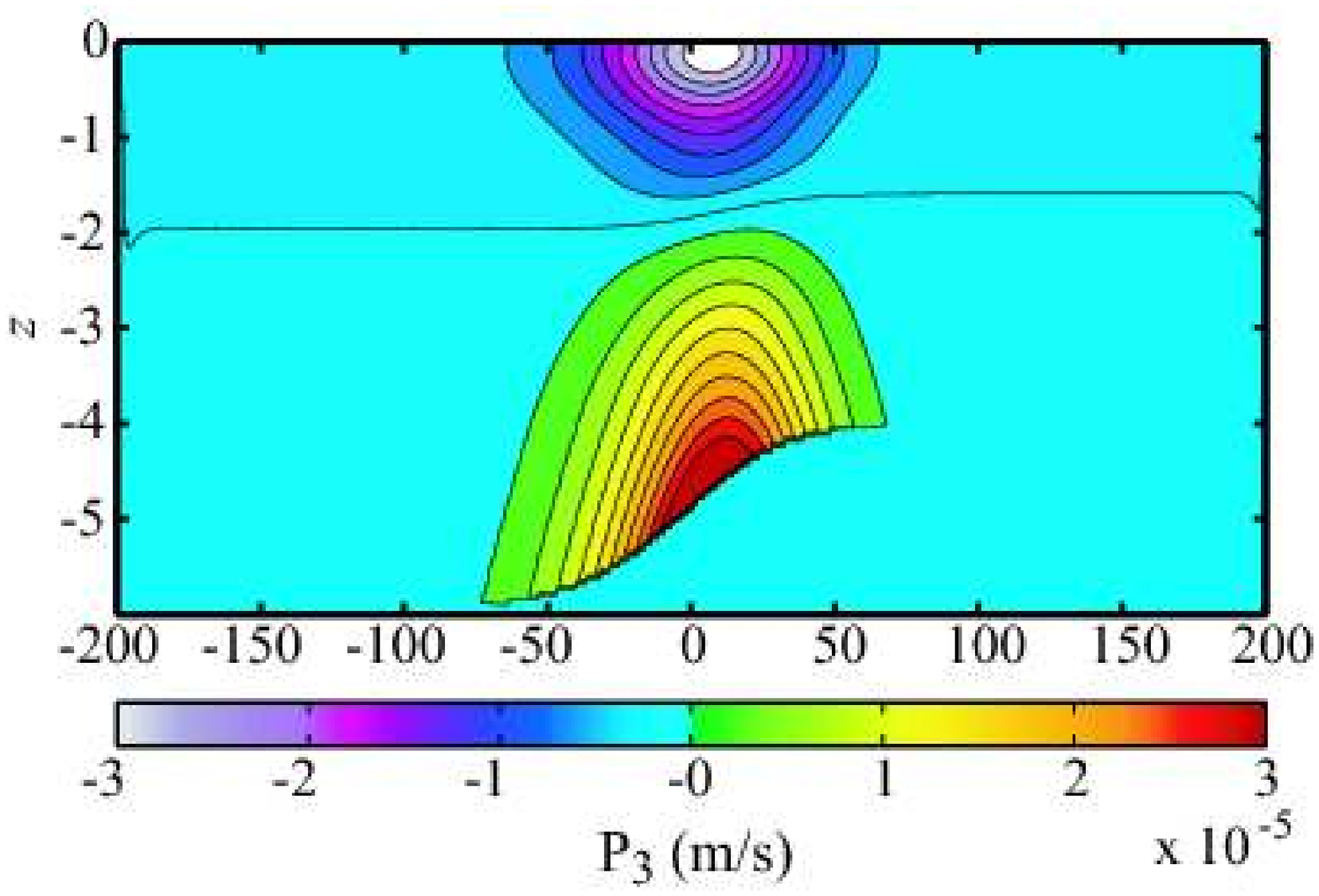}}
 \caption{Vertical wave pseudo-momentum for the same case as figure 2, estimated from eq. (\ref{P}), and verified to satisfy (\ref{P3}).}
 \label{figP3}
\end{figure}

For mild bottom slopes, the reflection coefficient is small as
predicted by Roseau (1976). The NTUA-nl2 model used here generally
gives accurate reflection coefficients, but it tends to
overestimate very weak reflections. In the first case investigated
here, the numerical reflection is $R=1\times 10^{-3}$, with no
significant effect on the wave dynamics. The NTUA-nl2 model is
used to provide the Fourier amplitudes of the mean, first and
second harmonic components of the velocity potential, over a grid
of 401 (horizontal) by 101 (vertical) points. From these
discretized potential fields, the mean, first and second harmonic
velocity components are obtained using second order centered
finite differences. As expected, the numerical solution gives a
horizontal mean flow $\overline{u}$ that compensates the
divergence of the wave mass transport and is thus of order
$\sigma/k \varepsilon^2$. Further $\overline{u}$ is almost uniform
over the vertical and is irrotational (figure \ref{figNTUA1}.b).
The vertical mean velocity is of higher order. The GLM momentum
balance is thus dominated by the hydrostatic and dynamic pressure
terms $p^H$ and $S^{\mathrm J}$. Although these two terms are
individually of the order of $0.01$~m$^2$~s$^{-2}$, their sum is
less than $2\times 10 ^{-16}$~m$^2$~s$^{-2}$ in the entire domain,
at the roundoff error level. It thus appears that this part of the
momentum balance is much more accurate than expected from the
asymptotic expansion. Indeed, for any bottom slope, in the limit
of small surface slopes and for irrotational flow and periodic
waves, the Stokes correction (\ref{Stokes correction}) for the
pressure and the time average of the Bernoulli equation give the
following expression for the modified kinematic pressure
(\ref{phat})
\begin{eqnarray}
    \widehat{p}=\frac{\overline{p}^L}{\rho_w}-\frac{\overline{u^l_j u^l_j}}{2}&=&\frac{\overline{p}
}{\rho_w}    +\frac{1}{\rho_w} \overline{\xi_j \frac{\partial \widetilde{p}}{\partial x_j}}-\frac{\overline{\widetilde{u}_j \widetilde{u}_j}}{2} \nonumber \\
     &=&-g z+\frac{1}{\rho_w}\overline{\xi_j \frac{\partial \widetilde{p}}{\partial x_j}}-
     \overline{\widetilde{u}_j \widetilde{u}_j }=-g z-\overline{\xi_j \frac{\partial^2 \widetilde{\phi}}{\partial x_j \partial t}}-
     \overline{\frac{\partial \widetilde{\xi_j}}{\partial t} \frac{\partial \widetilde{\phi}}{\partial x_j}}\nonumber \\
     &=&-g z-\frac{\partial}{\partial t}\overline{\xi_j
     \widetilde{u}_j}=-g z
\end{eqnarray}
where the equalities only hold to second order in the surface slope. Thus
 the kinematic modified pressure $\widehat{p}$ has no dynamical effect to
 second order in the wave slope, as already discussed by McWilliams et al.
 (2004) and Lane et al. (2007). For irrotational flow, this remains true for any bottom topography and
 even for rapidly varying wave amplitudes, including variations on scales
 shorter than the wavelength.

Thus the only wave effect is the static change in mean water level
(set-up or set-down), and dynamic consequences in the WBBL, where
$S^{\mathrm J}$ goes to zero, leaving the hydrostatic pressure
gradient to drive a mean flow that can only be balanced by bottom
friction. For slowly varying wave amplitudes the mean sea level is
given by Longuet-Higgins (1967\nocite{Longuet-Higgins1967}, eq.
F1)
\begin{equation}
\zb(x)= - \frac{k E}{\sinh (2 k D)}+  \frac{k_0 E_0}{\sinh (2 k_0)}
\label{zbF1}
\end{equation}
where the $0$ subscript correspond to quantities evaluated at any fixed
horizontal position, the choice of which being irrelevant to the estimation of
horizontal gradients of $\zb$.

\begin{figure}[htb]
 \vspace{9pt}
\centerline{\includegraphics[width=\textwidth]{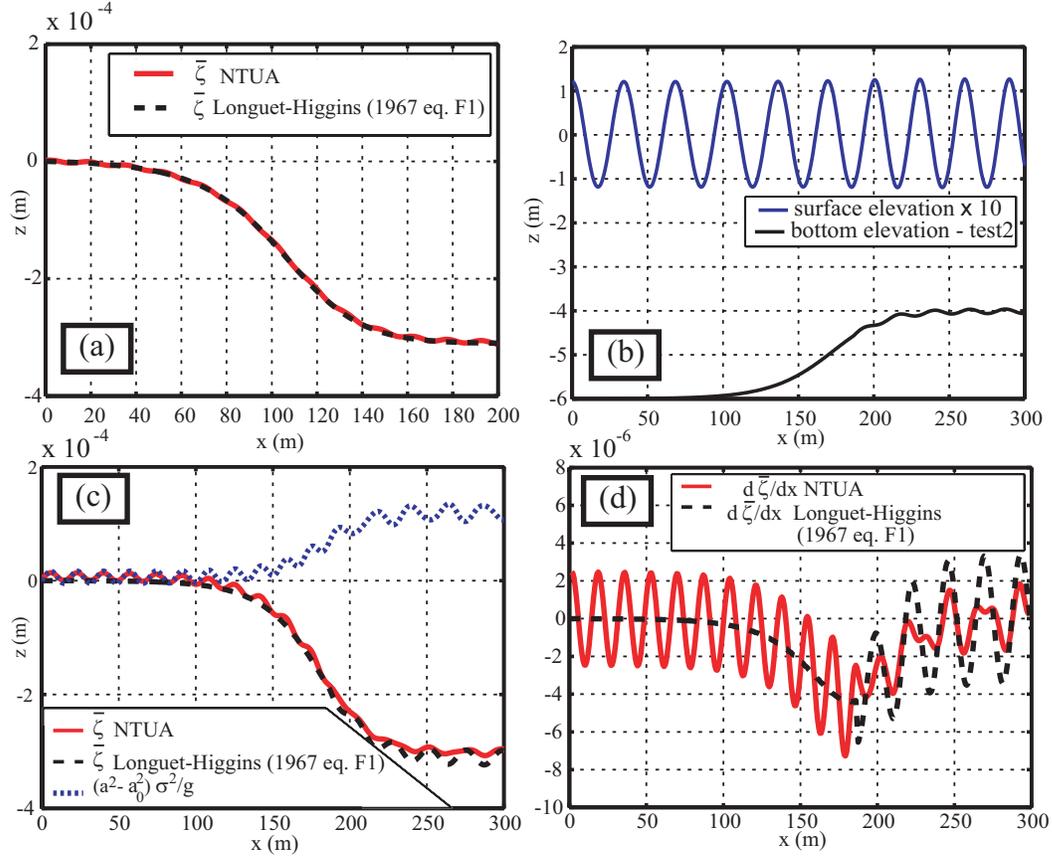}}
 \caption{(a) Mean sea level obtained with the NTUA-nl2 model (Belibassakis and Athanassoulis 2002) and the theory of Longuet-Higgins (1967 eq. F1: without
 standing waves) using conservation of the
 wave energy flux along the profile. (b) modified bottom profile resulting in a 3\% amplitude reflection at $\omega=1.2$ rad~s$^{-1}$,
(c) resulting mean sea level and normalized local wave amplitude $a$, (d) mean
sea level gradient (d).}
 \label{figNTUA2}
\end{figure}

Equation (\ref{zbF1}) is well verified by the NTUA-nl2 result for
the case considered so far (figure \ref{figNTUA2}.a). However,
this is no longueur true for rapid variations in the wave
amplitude $a(x)$, i.e. due to partially standing waves. In that
case one should use Longuet-Higgins' eq. D (op. cit.)
\begin{equation}
\zb(x)= - \left[\frac{\overline{\widetilde{u}_\beta \widetilde{u}_\beta -
\widetilde{u}_3 ^2}}{2g}\right]_{z=0} +
\left[\frac{\overline{\widetilde{u}_\beta \widetilde{u}_\beta - \widetilde{u}_3
^2}}{2g}\right]_{z=0,x=x_0}, \label{LHD}
\end{equation}
with $\widetilde{u}_\beta$ and $\widetilde{u}_3$ given by linear wave theory.
Eq. (\ref{LHD}) is a generalization of Miche's (1944a\nocite{Miche1944c}) mean
sea level solution under standing waves. Contrary to propagating wave groups,
for which the mean sea level is depressed under large waves, here the
depression occurs at the nodes of the standing wave, where the horizontal
velocities are largest and amplitudes are smallest (figure  \ref{figNTUA2}.c).

Eq. (\ref{LHD}) is well verified in the presence of partially
standing waves. To illustrate this, we have modified the bottom
topography, adding a sinusoidal bottom perturbation for $x>180$ m
with an amplitude of 5~cm and a bottom wavelength half of the
local waves' wavelength, which maximizes wave reflection (Kreisel
1949\nocite{Kreisel1949}). This yields a wave amplitude reflection
$R=0.03$, for $\omega=1.2$ rad~s$^{-1}$, of the order of observed
wave reflections over gently sloping beaches (e.g. Elgar et al.
1994\nocite{Elgar&al.1994}). The bottom is shown on figure
\ref{figNTUA2}.b. Although the standing wave pattern is hardly
noticeable in the surface elevation (the amplitude modulation is
only 6\%, figure \ref{figNTUA2}.c), the small pressure modulation
occur at much smaller scales, so that the associated gradient can
overcome the large scale gradients of the hydrostatic pressure
(figure \ref{figNTUA2}.d). As a result small partial stading waves
can dominating the momentum balance in the WBBL (see
Longuet-Higgins 1953, Yu and Mei 2000 for solutions obtained with
constant viscosity).

In the presence of such standing waves, and in the absence of
strong wave dissipation, the hydrostatic pressure on the scale of
the standing waves (e.g. given by Miche 1944a\nocite{Miche1944c})
drives the flow in the WBBL towards the nodes of the standing wave
(Longuet-Higgins 1953\nocite{Longuet-Higgins1953}), and is
balanced by bottom friction. This WBBL flow drives an opposite
flow above, closing a secondary circulation cell. This secondary
circulation is important for nearshore sediment transport just
outside of the surf zone (Yu and Mei 2000\nocite{Yu&Mei2000}). If
these sub-wavelength circulations are to be modelled, the present
$glm2z$-RANS theory should be extended to resolve the momentum
balance on the scale of partial standing waves.

 This extension is relatively simple as it only introduces
additional standing wave terms in all quadratic wave-related
quantities, arising from phase-couplings of the incident and
reflected waves. This extension provides a generalization of eq.
(\ref{LHD}) in the presence of other processes. For example,  eq.
(\ref{SJ3D}) now becomes
\begin{equation}
S^{\mathrm{J}}= g \int_{\mathbf{k_I}} \frac{k E({\mathbf
k})}{\sinh 2kD} \left[\left(1+R^2\right)- 2 R^2(\mathbf{k}) \cos(2
\psi'(\mathbf{k})) \right]
 {\mathrm
d}{\mathbf{k}} \label{SJ3Ds}
\end{equation}
with $R(\mathbf{k})$ the amplitude reflection coefficient and $2
\psi'(\mathbf{k})$ is the phase of the partial standing waves
defined by $\bnabla \psi'=\kb$ and $\partial \psi'/\partial t=-\kb
\bcdot \Ub_A t$  such that it is zero at the crest of the incident
waves. Note that the integral is over the incident wave numbers
only (e.g. for wave propagation directions from 0 to $\pi$).
Similar expressions are easily derived for the other wave forcing
terms.

\subsection{Effects of wave non-linearity}\label{Miche_section} Deep or intermediate water waves do not break very
often in most conditions (e.g. Banner et al. 2000, Babanin et al.
2001\nocite{Banner&al.2000,Babanin&al.2001}), thus the particular
kinematics of breaking or very steep waves likely contributes
little to the average forcing of the current. However, most of the
waves break in the surf zone and deviations from Airy wave
kinematics may introduce a systematic bias when the $glm2z$-RANS
equations are applied in that context. Many wave theories have
been developed that are generally more accurate than the Airy wave
theory (e.g. Dean 1970\nocite{Dean1970}). However, they may lack
some realistic features found in breaking waves, such as sharp
crests. In order to explore the magnitude of this bias, we shall
use the kinematics of two-dimensional incipient breaking waves as
given by the approximate theory of Miche
(1944b\nocite{Miche1944d}).

\begin{figure}[htb]
\label{test1}
 \vspace{9pt}
\centerline{\includegraphics[width=\textwidth]{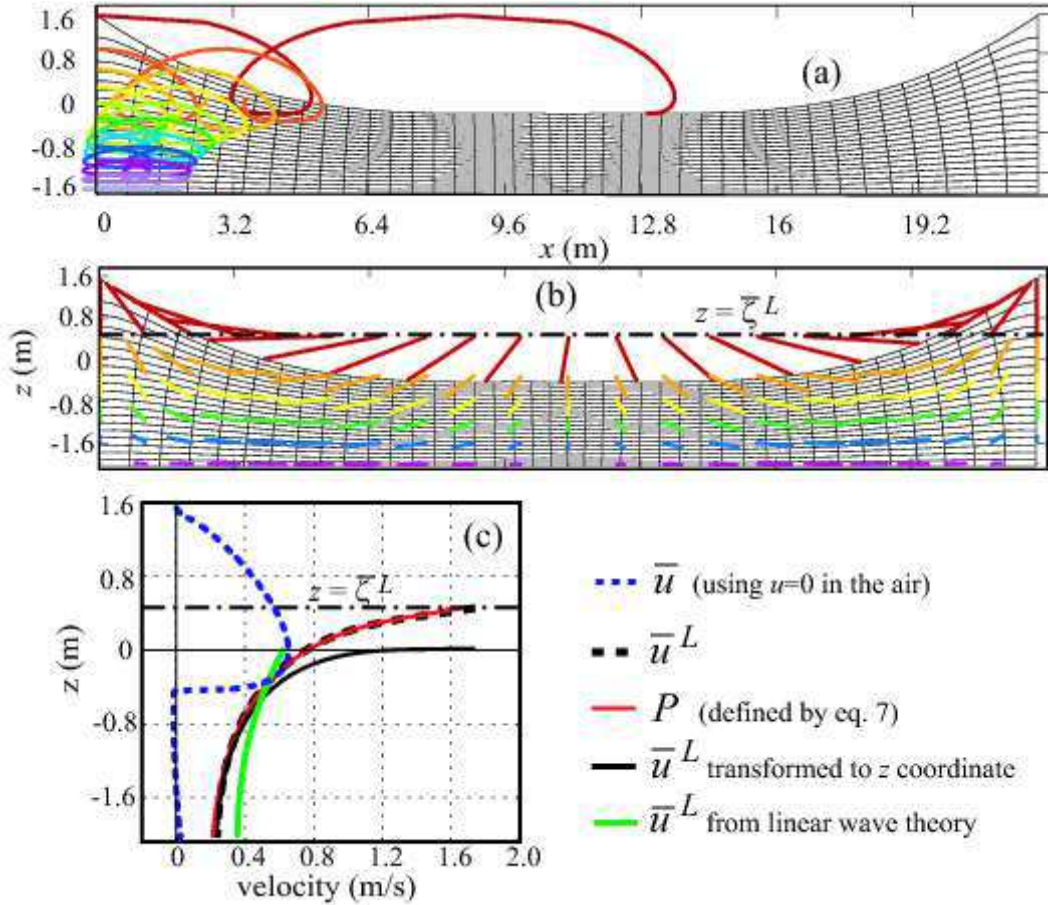}}
 \caption{(a) Illustration of the drift over 2 Eulerian periods in periodic Miche
 waves. Trajectories are color-coded with their initial depth, below a wave
 crest. The thin black lines are the lines of constant potential and
 streamfunction at $t=0$. (b) Field of displacements defining the GLM, as in figure 1.c. The dash-dotted line is the GLM position of the free surface $\overline{\zeta}^L$. (c) Profiles of Eulerian and Lagrangian mass transport velocity in a Miche wave compared to a linear
 wave with the same values of $k$ and $h$.}\label{Michefig}
\end{figure}

Miche's theory is based on the asymptotic expansion of the
potential flow from the triangular crest of a steady breaking
wave, extending Stokes' $120^\circ$ corner flow to finite depth.
From this Miche obtained his criterion for the maximum steepness
of a steady breaking wave, i.e. $h/\lambda=0.14 \tanh(kh)$ with
$h$ the breaking wave height and $\lambda$ the wavelength, which
favorably compares with observations. The Miche wave potential
$\phi$ and streamfunction $\widetilde{\psi}$ are expressed
implicitly as a function $G$ of the coordinates $x-x_c+\ir
(z-z_c)$, with origin on the wave crest $(x_c,z_c)$. The
coefficients in the series representing the reciprocal function
$G'$ are obtained from the boundary condition at the surface and
bottom. Unfortunately, these are imposed only under the wave crest
and trough, so that the bottom streamline may not be horizontal
away from the crest. This is particularly true for small values of
$kh$. Due to the expansion of $G'$ in powers of $\phi+\ir
\widetilde{\psi}$, the shape of the wave is nevertheless accurate
near the crest, and since the overall drift velocities are
dominated by the corner flow near the crest (see also
Longuet-Higgins 1979\nocite{Longuet-Higgins1979}), the
approximations of Miche have little consequence on the drift
velocities. The function $G'$ was modified here to make the bottom
actually flat, and the vertical under the trough an equipotential.
This deformation adds a weak rotational component to the motion
and the wave streamlines are weakly modified at the bottom under
the wave trough\footnote{This correction leads to negligible
differences compared to the exact solution as verified with
streamfunction theory to 60th order.}. The resulting wave for
$kh=0.58$ (corresponding to $b=1$ in Miche 1944b) is shown in
figure \ref{Michefig}.a. A numerical evaluation of that solution
is obtained at 201 equally spaced values of $\psi$  and 401
equally spaced values of $\phi$ (figure \ref{Michefig}.b). The GLM
displacement field $\xi$ is computed as described in section 2.1.
Since the streamlines are known in the frame of reference of the
wave, Lagrangian positions of 201 particles initially placed below
the crest at $x_i(0)=0$, were tracked over four Eulerian wave
periods. The positions $(x_i(t),z_i(t))$ are given by the
potential $\phi_i(t)$ and streamfunction $\psi_i$. The Lagrangian
period for each particle $T^L_i$ is determined by detecting the
first time when the particles pass under the crest again. The
Lagrangian mean velocity of each particle is then
$x_i(T^L_i)/T^L_i$, and it corresponds to a vertical position
$\overline{z}_i=\int_0^{T^L_i} z_i(t) \dr t$. This defines the
Lagrangian mean velocity $\overline{u}^L(\overline{z}_i)$ in GLM
coordinates. Following the coordinate transformation in section 2,
we further transform the GLM velocity profile to $z$ coordinate
(figure \ref{Michefig}.c). The resulting profile of
$\overline{\ub}^L$ has a horizontal tangent at $z=0$, as discussed
by Miche (1944b).

Contrary to Miche (1944b) who defined the phase speed $C$ of his wave by
imposing a zero mass transport, we have defined $C$ so that
$\Pb=\overline{\ub}^L$ with the pseudo-momentum $\Pb$ estimated from eq. (7)
using finite differences applied to the displacement field. The two profiles of
$\Pb$, estimated from eq. (7), and $\overline{\ub}^L$, estimated by time
integration of particle positions coincide almost perfectly. Thus the
estimation of $\Pb$ provides a practical method for separating the mean current
from the wave motion. Starting from any value of $C$, the difference between
$\overline{\ub}^L$ and $\Pb$ is the mean current velocity $\widehat{\ub}$. Here
$C$ was corrected to have $\widehat{\ub}=0$.

From $\xi$,  Bernoulli's equation can be used to obtain the GLM of
velocities and pressure. Compared to linear wave theory, the
Stokes drift in a Miche wave is much more sheared. It should be
noted that in the cnoidal theory investigated by Wiegel
(1959\nocite{Wiegel1959}) this drift velocity is depth-uniform.
Thus cnoidal wave theories may produce inaccurate results for 3D
wave-current interactions when extrapolated to breaking waves.
This marked difference in the 3D mean flow forcing due to breaking
waves compared to linear waves calls for a deeper investigation of
this question. Investigating such kinematics,
 may provide a rationale for the parameterization of
nonlinearity in the $glm2z$-RANS equations proposed here. Such a
parameterization is proposed by Rascle and Ardhuin (manuscript in
preparation for the Journal of Geophysical Research).

\section{Conclusion}
We have approximated the exact Generalized Lagrangian Mean (GLM)
wave-averaged momentum equations  of Andrews and McIntyre (1978a),
to second order in the wave slope, allowing for strong and sheared
mean currents with limited curvature in the current profile. These
approximated equations were then transformed by a change of the
vertical coordinate, giving a non-divergent GLM flow in $z$
coordinates. The resulting conservation equations for horizontal
momentum (\ref{GLM2z}) and mass (\ref{GLMmasszb}), with boundary
conditions (\ref{GLMsfcbc})--(\ref{BBL1b}) may be solved using
slightly modified versions of existing primitive equations models,
forced with the results of spectral wave models. Although the
Stokes drift introduces a source of mass at the surface for the
quasi-Eulerian flow, this is does not pose any particular problem,
and such mass source have long been introduced for the simulation
of upwellings. The HYCOM model (Bleck 2002\nocite{Bleck2002}) was
modified by R. Baraille to solve a simplified set of the present
equations, retaining only the wave-induced mass transport in both
the mass and momentum equations, and the tracer equation (in which
the advection velocity is simply $\overline{\ub}^L$, see also
MRL04). This work was applied to the a hindcast of the
trajectories of sub-surface oil pellets released by the tanker
Prestige-Nassau, which sank off Northwest Spain in November 2002
(presentation at the 2004 WMO-JCOMM `Oceanops' conference held in
Toulouse, France). The full equations derived here have also been
implemented in the ocean circulation model ROMS (Shchepetkin and
McWilliams 2003\nocite{Shchepetkin&McWilliams2003}), and results
will be reported elsewhere. The equations presented here have also
been applied for the modelling of the ocean mixed layer in
horizontally-uniform conditions (Rascle et al. 2006).

Although a general expression for the turbulent closure has been
given, it has not been made explicit in terms of the wave and mean
flow quantities beyond a heuristic closure that combines an eddy
viscosity mixing term with the known sources of momentum due to
wave dissipation. A proper turbulent closure is left for further
work, possibly extending and combining the approaches of Groeneweg
and Klopman (1998), with those of Teixeira and Belcher (2002).
Further, some wave forcing quantities have been expressed in terms
of the Eulerian mean current $\overline{\ub}$ instead of the
quasi-Eulerian mean current $\widehat{\ub}$. The conversion from
one to the other, can be done using eq. (\ref{Eul-quasi}), to the
order of approximation used here. However, it would be more
appropriate, in particular for large current shears, to start from
quasi-Eulerian wave kinematics, instead of Eulerian solutions of
the kind given by Kirby and Chen (1989, our eq. 10--12).

Beyond the turbulence closure, there are essentially two practical
limitations to the approximate $glm2z$-RANS equations derived
here. First, the expansion of wave quantities to second order in
the surface slope is only qualitative in the surf zone. Although
this was acceptable in two dimensions (see Bowen
1969\nocite{Bowen1969} and most of the literature on this
subject), it is expected to be insufficient in three dimensions
due to a significant difference in the profile of the wave-induced
drift velocity $\Pb$, which exhibits a vertical variation with
surface values exceeding bottom values by a factor of 3, even for
$kh<0.2$ in which case linear wave theory predicts a depth-uniform
$\Pb$. This conclusion is based on both the approximate theory of
Miche (1944b), and results of the streamfunction theory of
Dalrymple (1974) to 80th order\nocite{Dalrymple1974}. Such
numerical results can be used to provide a parameterization of
these effects. Further investigations using more realistic
depictions of the kinematics of breaking waves will be needed.
Second, the vertical profile of the mean current in the surf zone
may be such that the wave kinematics are not well described by the
approximations used here. A strong nonlinearity combined with a
strong current shear and curvature can lead to markedly different
wave kinematics (e.g. da Silva and Peregrine
1988\nocite{TelesdaSilva&Peregrine1988}).

With these caveats, the equations derived here provide a generalization of
existing equations, extending Smith (2006) to three dimensions and vertically
sheared currents, or McWilliams et al. (2004) to strong currents. Of course,
mean flow equations can be obtained, at least numerically, using any solution
for the wave kinematics with the original exact GLM equations, as illustrated
in section~\ref{Miche_section}. The wave-forcing on the mean flow is a vortex
force plus a modified pressure, a decomposition that allows a clearer
understanding of the wave-current interactions, compared to the more
traditional radiation stress form. This is most important for the
three-dimensional momentum balance and/or in the presence of strong currents,
e.g. when a rip current is widened by opposing waves, as observed by Ismail and
Wiegel (1983\nocite{Ismail&Wiegel1983}) in the laboratory. Such a situation was
also recently modelled by Shi et al. (2006\nocite{Shi&al.2006}).

{\it Acknowledgments.} The correct interpretation of the vertical
wave pseudo-momentum $P_3$ would not have been possible without
the insistent questioning of John Allen.  The critiques and
comments from  Jaak Monbaliu and Rodolfo Bola{\~n}os helped
correct some misinterpretation of the equations and greatly
improved the present paper. N.R. acknowledges the support of a
CNRS-DGA doctoral research grant.

\bibliographystyle{elsart-harv}
\bibliography{../references/wave}
\setcounter{table}{1}
\begin{table}
  \centering
  \begin{tabular}{ccc}
\hline
  Symbol       & name  & where defined  \\
\hline
  1 and 2&  indices of the horizontal dimensions               &  after (\ref{dispersion}) \\
  $3$&  index of the vertical dimension               &  after (\ref{dispersion})  \\
  $a$&   wave amplitude               &  after (\ref{w}) \\
  $D=h+\zb$&   mean water depth               &   after (\ref{P}) \\
  $\mathbf{f}=(f_1,f_2,f_3)$& Coriolis parameter vector (twice the rotation vector)              & after (\ref{Prot})   \\
  $F_{CC}$, $F_{CS}$, $F_{SC}$ and $F_{SS}$&Vertical profile functions      &  after (\ref{w})  \\
  $g$ & acceleration due to gravity and Earth rotation & after (\ref{P}) \\
  $h$&   depth of the bottom (bottom elevation is $z=-h$)              &  before (\ref{dispersion}) \\
  $J$&  Jacobian of GLM average               &  after (\ref{GLMmass}) \\
  $\kb=(k_1,k_2)$&   wavenumber vector                &  after (\ref{P}) \\
  $K_1$& Depth-integrated vertical vortex force                & (\ref{K1}) \\
  $K_2$& Shear-induced correction to Bernoulli head                & (\ref{K2p}) \\
  $K_z$& vertical eddy viscosity              & (\ref{closure_param}) \\
$(\bcdot)^l$  & Lagrangian perturbation  & (\ref{Lagr.pert}) \\
$\overline{(\bcdot)}^L$  & Lagrangian mean  & (\ref{phiL}) \\
$m$  & shear correction parameter & (\ref{m}) \\
$\mathbf{M}$  & depth-integrated momentum vector & (\ref{Mtot}) \\
$\mathbf{M}^w$  & depth-integrated wave pseudo-momentum vector & (\ref{Mw}) \\
$\mathbf{M}^m$  & depth-integrated mean flow momentum vector & after (\ref{Mw}) \\
$\mathbf{n}$  & unit normal vector & (\ref{n}) \\
\hline
\end{tabular}
  \caption{Table of symbols}\label{table_symb}
\end{table}
\setcounter{table}{1}
\begin{table}
  \centering
  \begin{tabular}{ccc}
\hline
  Symbol       & name  & where defined  \\
\hline
$p$  & full dynamic pressure & after (\ref{GLM_Gro}) \\
$\widetilde{p}$  & wave-induced pressure &  (\ref{p}) \\
$p^H$ & hydrostatic pressure & after (\ref{pbar2L}) \\
$\Pb=(P_1,P_2,P_3)$&   wave pseudo-momentum                & (\ref{Prot})   \\
$t$                      & time               & before (\ref{phiL})  \\
$\ub=(u_1,u_2,u_3)$      & velocity vector                &   \\
$\widetilde{\ub}$        & wave-induced velocity                & (\ref{u1}) and (\ref{ubl})  \\
$\overline{\mathbf{u}}^L$& Lagrangian mean velocity                 & after (\ref{phiL}) \\
${\mathbf u}_{A}$       & advection velocity for the wave action & (\ref{uA}) \\
$\hu_\alpha=\uL_\alpha-P_\alpha$     & quasi-Eulerian horizontal velocity & before (\ref{Eul-quasi}) \\
$s=z+\overline{\xi}_3^L$ & GLM to $z$ transformation function                 & (\ref{sdef}) \\
$\overline{(\bcdot)}^S$  & Stokes correction                 & (\ref{Stokes correction}) \\
$S_{ij}$&  stress tensor                  & (\ref{viscstress}) \\
$S^{\mathrm J}$& wave-induced kinematic pressure               & (\ref{SJ3D}) \\
$S^{\mathrm{Shear}}$& shear-induced correction to $S^{\mathrm J}$            & (\ref{Sshear}) \\
$w=u_3$& vertical velocity               &  before (\ref{GLM_vert})  \\
$\hw=\uL_3-P_3$     & quasi-Eulerian vertical velocity & before (\ref{GLM_vert}) \\
$W$& GLM vertical velocity in $z$ coordinates             & (\ref{DefOmega})  \\
$\xb=(x_1,x_2,x_3)$ &   position vector                &   before (\ref{phiL}) \\
\hline
\end{tabular}
\nonumber   \caption{Table of symbols, continued}\label{table_symb2}
\end{table}
\setcounter{table}{1}
\begin{table}
  \centering
  \begin{tabular}{ccc}
\hline
  Symbol       & name  & where defined  \\
\hline
$\mathbf{X}$ & diabatic source of momentum & after (\ref{Eul-quasi}) \\
  $\widehat{\mathbf{X}}$ & diabatic source of quasi-Eulerian mean momentum &
  (\ref{Turb_closure}) \\
  $z=x_3$& vertical position               &  after (\ref{dispersion})   \\
  $\alpha$ and $\beta$&  dummy indices for horizontal dimensions     &  \\
  $\delta_{ij}$  & Kronecker's symbol, zero unless $i=j$     &  after (\ref{GLM_Gro}) \\
  $\varepsilon$  &  generic small parameter     & after  (\ref{dispersion}) \\
  $\varepsilon_1$  &  maximum wave slope     & after   (\ref{P})  \\
  $\varepsilon_2$  &  maximum horizontal gradient parameter     & after   (\ref{P})  \\
  $\varepsilon_3$  &  maximum current curvature parameter    &   (\ref{eps3})  \\
  $\epsilon_{i j k} A_j B_k$  & component $i$ of the vector product ${\mathbf A}\times {\mathbf B}$    &   after (\ref{Prot})  \\
  $\zeta$& free surface elevation                 & before  (\ref{dispersion})  \\
  $\lambda$& wavelength               &  section 4.2 \\
  $\nu$& kinematic viscosity of water                 &  after (\ref{viscstress}) \\
  $\xi=(\xi_1,\xi_2,\xi_3)$& wave-induced displacement                & before (\ref{phiL})  \\
  $\rho_w$&  density of water (constant)                 &  after (\ref{w})  \\
  $\sigma$&  relative radian frequency                  & after (\ref{P}) \\
  $\tau_{ij}$&  mean stress tensor                  & (\ref{taucontsurf}) \\
  $\psi$&   wave phase                  &after (\ref{P}) \\
  $\omega$&   absolute radian frequency                  & after (\ref{P})  and (\ref{dispersion}) \\
  $\Omega_3$&  depth-weighted vertical vorticity of the mean flow               &  (\ref{Omega_3}) \\
  $\bnabla$&   horizontal gradient operator                 & after (\ref{P}) \\
\hline
\end{tabular}
 \nonumber \caption{Table of symbols, continued}\label{table_symb3}
\end{table}

\end{document}